# Spitzer Evidence for a Late Heavy Bombardment and the Formation of Urelites in η Corvi at ~1 Gyr

**C.M. Lisse[1], M. C. Wyatt[2], C. H. Chen[3], A. Morlok[4], D.M. Watson[5], P. Manoj[5], P. Sheehan[5], T. M. Currie[6], P. Thebault[7], and M.L. Sitko[8]**

[1]JHU-APL, 11100 Johns Hopkins Road, Laurel, MD 20723 carey.lisse@jhuapl.edu.

[2]Institute of Astronomy, University of Cambridge, Madingley Road, Cambridge CB3 0HA, UK wyatt@ast.cam.ac.uk.

[3]STScI 3700 San Martin Dr. Baltimore, MD 21218 cchen@stsci.edu.

[4] Dept. of Earth and Planetary Sciences, The Open University, Milton-Keynes, UK a.morlok@open.ac.uk

[5]Department of Physics and Astronomy, University of Rochester, Rochester, NY dmw@pas.rochester.edu, manoj@pas.rochester.edu, psheeha2@mail.rochester.edu

[6]Code 667, NASA-GSFC, Greenbelt, MD 20771 thayne.m.currie@nasa.gov

[7]Observatoire de Paris, F-92195 Meudon Principal Cedex, France, philippe.thebault@obspm.fr

[8]Space Science Institute, 475Walnut St., Suite205, Boulder, CO80301 sitko@spacescience.org

Proposed Running Title: **Warm KBO Dust in η Corvi**

Please address all future correspondence, reviews, proofs, etc. to:

Dr. Carey M. Lisse

Planetary Exploration Group, Space Department

Johns Hopkins University, Applied Physics Laboratory

11100 Johns Hopkins Rd

Laurel, MD 20723

240-228-0535 (office) / 240-228-8939 (fax)

Carey.Lisse@jhuapl.edu

# ABSTRACT

We have analyzed *Spitzer* and NASA/IRTF 2 – 35 μm spectra of the warm, ~350 K circumstellar dust around the nearby MS star *η Corvi* (F2V, 1.4 ± 0.3 Gyr). The spectra show clear evidence for warm, water- and carbon-rich dust at ~3 AU from the central star, in the system's Terrestrial Habitability Zone. Spectral features due to ultra-primitive cometary material were found, in addition to features due to impact produced silica and high temperature carbonaceous phases. At least 9 x $10^{18}$ kg of 0.1 – 100 μm warm dust is present in a collisional equilibrium distribution with dn/da ~ $a^{-3.5}$, the equivalent of a 130 km radius KBO of 1.0 g/$cm^3$ density and similar to recent estimates of the mass delivered to the Earth at 0.6 – 0.8 Gyr during the Late Heavy Bombardment. We conclude that the parent body was a Kuiper-Belt body or bodies which captured a large amount of early primitive material in the first Myrs of the system's lifetime and preserved it in deep freeze at ~150 AU. At ~1.4 Gyr they were prompted by dynamical stirring of their parent Kuiper Belt into spiraling into the inner system, eventually colliding at 5-10 km/sec with a rocky planetary body of mass ≤ $M_{Earth}$ at ~3 AU, delivering large amounts of water (>0.1% of $M_{Earth's\ Oceans}$) and carbon-rich material. The *Spitzer* spectrum also closely matches spectra reported for the Ureilite meteorites of the Sudan Almahata Sitta fall in 2008, suggesting that one of the Ureilite parent bodies was a KBO.

## 1. Introduction

Solar system chronology and *Spitzer* studies of debris disks provide a strong motivation for investigating the nature of collisions in Gyr-old star systems. The lunar and terrestrial impact records suggest a high frequency of collisions, tailing down rapidly thereafter, at 0.6—0.8 Gyr after the birth of our solar system (the so-called Late-Heavy Bombardment or LHB; Tera *et al.* 1974; Wetherill 1975; Kring & Cohen 2002; Ryder, 2002, 2003; Chapman *et al.* 2007; Jørgensen *et al.* 2009). Numerical models show that planetary migration of Jupiter and Saturn through a 2:1 orbital resonance at ~1 Gyr can force Uranus and Neptune to migrate outwards and destabilize an originally massive Kuiper belt planetesimal population, causing more than 99% of the planetesimals to dynamically scatter, eventually leaving the system, or, for a minority, colliding with another orbiting body (Gomes *et al.* 2005; Tsiganis *et al.* 2005). Using sub-mm imaging, Wyatt *et al.* 2005 detected a luminous, massive cold debris belt around *η* Corvi at ~1 Gyr, in an ~300 AU diameter ring tilted at about $45^o$ to the line of sight from Earth, suggesting that the system's Kuiper Belt is dynamically active and supplying the cold debris belt through collisions. IRAS measurements in 1983 of the same system showed a definite strong excess of emission over the stellar photosphere at 12 and 25 μm, demonstrating the presence of warm circumstellar dust within a few AU of the primary as well, dust which could have been created in an LHB-like collision. The timing, high dust luminosity, presence of both warm and cold dust, and structure in the cold dust belt thus make *η* Corvi a good candidate for a system that contains massive, migrating planets that are in the process of dynamically destabilizing their surrounding icy outer planetesimal population, while also causing a KBO – inner system object collision.

Optically-thin, gas-poor debris *disks* around 5 Myr to 5 Gyr-old stars are important signposts for planet formation and evolution, since their sustained emission implies the presence of a reservoir of massive, colliding self-gravitating planetesimals (e.g. Backman and Paresce 1993; Wyatt 2008; Kenyon and Bromley 2008). The prevalence of debris disks around sun-like stars and more massive stars (20 to 60%; Rieke *et al.* 2005; Su *et al.* 2006; Currie *et al.* 2008, Carpenter *et al.* 2009a) implies that the formation of these planetesimals is common; our own solar system contains debris dust and planetesimals in two small belts – warm dust and rocky, metal-rich bodies in the asteroid belt at 2 to 4 AU, cold dust and icy planetesimals in the Kuiper belt at 30 to 60 AU.

Therefore, constraining the evolution and composition of debris disks around other stars provides a crucial context for the formation and present-day properties of the solar system.

Large photometric studies of many debris disks and targeted spectroscopic studies of individual debris disks conducted with the *Spitzer Space Telescope* have constrained the typical locations of debris-producing collisions, the evolution of the emission from these collisions, and, in some cases, the chemical composition of and events identified by the colliding parent bodies. Most debris disks have dust temperatures colder than ~ 150—200 K, consistent with debris produced from icy colliding planetesimals (e.g. Chen *et al.* 2006; Carpenter *et al.* 2009b). The frequency of warm (> 200 K) debris dust is low, < 5—10% regardless of age, indicating that planetesimal collisions in the terrestrial zone or asteroid belt-like regions are rare or have a small observational window, consistent with expectations from theory (e.g. Carpenter *et al.* 2009b; Morales *et al.* 2009; Currie *et al.* 2007, 2008b; Kenyon and Bromley 2004; Wyatt *et al.* 2007; Moor *et al.* 2010).

With the possible exception of the youngest debris disks (~5—20 Myr), mid-to-far IR debris disk luminosities define an envelope that decays with time as ~ 1/t, consistent with a low-velocity (< 1 km/sec) collisional grinding of a planetesimal belt that has finished growing larger bodies, especially for Gyr-old systems (e.g. Rieke *et al.* 2005; Su *et al.* 2006; Wyatt 2008; Löhne *et al.* 2008; Currie *et al.* 2008, 2009; Kenyon and Bromley 2008, 2010; Hernandez *et al.* 2009). As a whole, most debris disk spectra are featureless, consistent with optically-thick grains (> 10 microns) and thus a lack of copious, small debris from massive collisions (e.g. Chen *et al.* 2006; Carpenter *et al.* 2009a). The rare debris disks with strong mid-IR spectral features typically have young ages (~ 10—20 Myr) and high luminosities, likely due to copious production of warm small dust by major processes of solar system formation and evolution, such as terrestrial planet formation, catastrophic impacts like the Moon-forming event, and terrestrial zone water delivery (e.g. Lisse *et al.* 2008, 2009; Currie *et al.* 2010).

In light of *Spitzer* photometric and spectroscopic studies, the large majority of debris disks around Gyr-old stars seem to almost always (1) have low luminosities consistent with material produced slowly by collisional grinding and parent body sublimation, (2) lack strong solid state features produced by copious fine, but ephemeral dust and (3) lack warm, inner system dust. Therefore, older debris disks with high luminosities, solid states spectral features, and warm dust comprise an exceptionally rare class of objects (frequency ~ 1-2%, Gaspar *et al.* 2009) that may be diagnostic of recent, massive collision events in the inner few AU of a system and may provide a context for important dynamical effects that occurred in the solar system well after planet formation finished. E.g., BD 20 +307 and HD 106797 are well-studied examples of Gyr-old stars surrounded by unusually luminous warm debris disks; spectral modeling reveals solid state spectral features produced by transient dust from recent collisional events (e.g. Song *et al.* 2005; Fujiwara *et al.* 2009; Weinberger *et al.* 2011).

*η* Corvi, a nearby main sequence star with a debris disk, is another rare candidate for study, as it contains detectable cold ***and*** warm dust reservoirs of high luminosity presenting strong spectral features. Of the 59 IRAS excess systems observed by *Spitzer* and studied by Chen *et al.* 2006, it is the 3rd brightest and exhibits the 3rd highest disk luminosity ($L_{IR}/L_* = 3 \times 10^{-4}$). Of 44 stellar systems now observed by the Keck Interferometer, it is one of only 3 for which an extended disk of warm dust emission at 10 μm was significantly detected (Stark *et al.* 2009, Millan-Gibet *et al.* 2009, 2011), at a level of 1389 ± 273 zodis (1σ). However, unlike most mid-IR high-luminosity debris disks, it is an old main sequence star (~ 1 Gyr), and its excess emission at 24 μm lies well outside the range defined by most stars of its age (Figure 1; Wyatt 2008). Furthermore, simple modeling of its optical through far-IR spectral energy distribution (SED; Figure 1), shows that *η* Corvi's dust emission plausibly originates from 2 greybody sources of temperatures ~35K and 350 K (Bryden *et al.* 2009). While the ~35K reservoir matches well with the predicted temperature of the cold dust at ~ 150 AU detected in the sub-mm by Wyatt *et al.* (2005), the temperature of the warm component implies the presence of copious dust in terrestrial zone/asteroid belt-like regions. Bryden *et al.* (2009) obtained interferometric measurements demonstrating that the middle regions of the system are empty of dust, and the system structure qualitatively consistent with dynamical sculpting of sub-mm emitting dust by ice/gas-giant planets (Wyatt *et al.* 2005, Kuchner & Stark 2010). Analysis of the *η* Corvi debris disk may thus provide an important study of LHB-like events which occurred well after a system's planets finished forming.

In this paper we analyze in detail *Spitzer η* Corvi IRS 5 to 35 μm and IRTF/SPeX 2 to 5 μm spectral measurements of the warm dust component, utilizing strong features in the near- and mid-IR spectrum to allow us to determine the major dust components producing the observed emission. We have used a similar methodology to study the amount, kind, and location of the dust and water gas/ice excavated from the comet 9P/Tempel 1 (Lisse *et al.* 2006), the dust and water ice found in comet C/1995 O1 (Hale-Bopp) and the comet dominated primordial disk around the Herbig A0 star HD100546 (Lisse *et al.* 2007a), in the dense zody cloud around the 2-10 Gyr old K0V star HD69830 (Lisse *et al.* 2007b), in near orbit around the ancient DZ white dwarf GD29-38 (Reach *et al.* 2008), and in the terrestrial planet forming region around the young A/F stars HD113766 (Lisse *et al.* 2008) and EF Cha (Currie *et al.* 2010). We were able to obtain a good model fit to the *η Corvi* warm dust spectrum and find that the best remote sensing spectral match is from the very primitive material found at 13 AU in the gas-rich, giant planet forming system HD100546, and the best solar systems material match is to the mixed polyglot primitive carbon rich/differentiated igneous materials found in the Sudan Almahata Sitta Ureilite meteor fall of 2008. The *η Corvi* warm disk material is even more primitive and carbon-rich than the material found in comets, with a total amount of mass that is found in large Centaur or medium sized KBO objects in our solar system. The best-fit particle size distribution (PSD) for the warm dust is close to that expected for a system in collisional equilibrium with the smallest particles removed by radiation pressure. These facts coupled together argue for a very primitive icy parent body, formed very early in the system's history and large enough to capture and gravitationally bind the most volatile of protoplanetary carbon species. Coupled with the ~$3 \times 10^4$ times more massive extended cold sub-mm disk of Wyatt *et al.*, we infer that the Kuiper Belt of the *η Corvi* system at ~1 Gyr is indeed in an excited state, leading to scattering of large primitive bodies into the inner regions of the system, where moderate velocity impacts and subsequent collisional grinding processes release their icy, carbon-rich dust, similar to the events predicted to have formed the Ureilite meteor parent body and to have occurred at 0.6 - 0.8 Gyr after CAI formation during the Late Heavy Bombardment (LHB) in our own Solar System.

## 2. Observations

**2.1 Stellar Data for the *η* Corvi system**. *η Corvi* (HR4775, HD109085), the 6$^{th}$ brightest star in the northern constellation Corvus, is a solitary F2V main sequence ("dwarf") star located 18.2 pc away from Earth, absolute $V_{mag}$ = 2.99, of approximately solar metallicity (Casagrande *et al.* 2011), and is somewhat bigger (M = 1.4 $M_{\odot}$) and hotter ($T_{eff}$ ~6800 K) and about 5 times more luminous than the Sun (Table 1). To 0$^{th}$ order, the BVR/2MASS/IRAS/*Spitzer* spectral energy distribution (SED; Figure 2) for the circumstellar material orbiting *η* Corvi is well described by a Kurucz F2V photospheric emission model plus the emission from 2 greybody sources at temperatures ~350 and 35 K with $L_{IR}/L_* = 3 \times 10^{-4}$ (Bryden *et al.* 2009). Based on its stellar type, rotation rate, metallicity, and x-ray activity, *η* Corvi's age was estimated by Nordstrom *et al.* (2004) to be 1.4 ± 0.3 Gyr; from its position relative to evolutionary tracks in the HR diagram, Mallik, Parthasarathy & Pati (2003) determined a stellar age of 1.3 Gyr. Mamajek (2011, private communication), using Bertelli *et al.* 2009 isochrones and evolutionary tracks and Z=0.017 and Y=0.26 for a solar composition star, finds an age of 1380 ± 190 Myr and a mass of 1.44 ± 0.01 $M_{\odot}$. In this work we adopt a working age of 1.4 Gyr for *η* Corvi in calculations, but noticing the range of age estimates and typical age uncertainties on the order of a factor of 2, also write the age as ~1 Gyr for the sake of argument.

**2.2 *Spitzer* and IRTF Observations of η Corvi.** For their IRS observations of η Corvi conducted on 05 Jan 2004, Chen *et al.* (2006) utilized a combination of the Short-Low (5.2–14.0 μm, λ/Δλ ~ 90), Short-High (9.9–19.6 μm, λ/Δλ ~ 600), and Long-High (18.7-37.2 μm, λ/Δλ ~ 600) modules in IRS mapping mode. In order to avoid time-consuming peak-up, the observatory was operated in IRS spectral mapping mode where a 2 × 3 raster (spatial × dispersion, centered on the nominal star position) was performed, with 1 cycle of 6 seconds integration at each position (Watson *et al.* 2004). Because of significant off-center pointing alignment in the as-performed raster map, and great improvements over the last 5 years in the understanding of IRS pointing effects, individual pixel gain, dark current, linearity, and bad pixel behavior, the spectra were re-extracted from the *Spitzer* archive for this work and re-calibrated using SMART Version 17 with optimized, model PSF based extraction for the SL order (Lebouteiller *et al.* 2010) and defringing/tilting of the LH orders (Higdon *et al.*, 2004). Correcting for the off-center pointing was found to be especially important, as it could have led to potential changes in response up to 50% in the short wavelength orders, which could have potentially been the cause of the interesting and complicated structure in the 5 – 9 um region of the η Corvi spectrum.

Re-reduction of the SL data was done using the opposite nod position for sky subtraction and Optimal Point Source extraction (OPS) in the SMART program. With OPS, the SL spectra is extracted by fitting an empirically derived point spread function to the profile of the object that optimizes the SNR of the observation in each wavelength bin. This is different from the standard

"tapered column" extraction method, which is based on co-adding the flux in the cross-dispersion direction within a window large enough to contain (most of) the source flux for a typical point source and results in much more stable measurements at the cost of greatly increased time and computing power expenditures per spectral extraction. The high resolution SH and LH data included no sky data and were extracted with a full slit extraction. The local background was determined using the signal level in the wings of the PSF. The LH data also suffered from "order tilting" or "bowing" of the spectral orders, seen as mismatches between orders and non-physical tilting of the continua as a result of a time- and position-dependent current on the array. To correct this we employed the *Spitzer* Science Center program ***DarkSettle*** (Ref: *Spitzer* Data Analysis Cookbook) which uses a set of BCD data to calculate a time- and position-dependent smoothed background and removes this from the data. Data from standard stars α Lac for SL and ξ Dra for SH and de-tilted LH were prepared identically, and then template spectra were divided by the data to produce Relative Spectral Response Functions (RSRFs). Our η Corvi data was then multiplied by these RSRFs to calibrate the flux.

For this work we have used SL data from 5.2 to 9.9 um, SH data from 9.9 to 19 um, and LH data from 19 to 35 um. Care was taken to verify the robustness of the derived spectra; e.g., the spectra were extracted by two different analysts, and the results were found to agree closely, within ±2σ, for each spectral bin. We have also checked to verify that the spectral features in the SL module were found in both nods, and that they disappear in the difference to better than 2σ, arguing that they are due to real on-sky photoresponse and not detector systematics. Comparison of the extracted SL and SH fluxes in the 9.9 – 14.3 um region where they overlap, produced excellent agreement, within 2σ, verifying that the relative calibration used for the SL and SH orders was good; MIPS and IRS photometry at 16, 22, and 24 μm was used to tie down the longer wavelength absolute calibration of the SH and LH orders to within 10%. In the end, a total of 1764 independent spectral points were obtained over the range 5.2 to 37 μm. The resulting spectra have 117 data points taken with the SL orders at R = 60 to 120 from 5.2 to 9.9 um, and 1647 data points taken with the SH/LH orders at R = 600 to 650 from 9.9 to 35 um. The median SNR of the SL data is 24, and of the SH/LH data is 37.

Supporting observations of *η Corvi* at 2.2 to 5.1 μm were made by our group on 21 Apr 2011 UT using the SPeX near infrared spectrometer (Rayner *et al.* 2003, Vacca *et al.* 2004) on the NASA/IRTF 3m telescope on Mauna Kea. The data were obtained in LXD 2.1 mode at R ~ 2500 with the narrow 0.3" slit (5.4 AU at 18 pc) centered on the star using 300 sec integration times and on-sky nodding. The airmass = 1.47, SNR > 80 *η Corvi* observations were interleaved with observations of nearby A-star standards and reduced using the SPeXTools software package (Cushing *et al.* 2003) in a manner identical to that used for the 200 stars of the SPeX Cool Star library (Rayner *et al.* 2009). The resulting stellar spectra, demonstrating good H-atom emission lines, compared well to dwarf F-star standards listed in the spectral library and Kurucz models of an F2V stellar photosphere. Due to opacity issues, only the high fidelity 2.2 – 2.5, 2.8 - 4.2, and 4.5 – 5.1 μm data from the high atmospheric transmission windows were used in this work.

The *η Corvi* disk excess flux was then calculated by removing the stellar photospheric contribution from the IRS and SPeX spectra. The photospheric contribution was modeled by assuming that the *η Corvi* spectrum is represented by an F2V star with an age of 1.4 ± 0.3 Gyr (Norstrom *et al.* 2004). The star was assumed to have solar abundance, log g = 3.9, and E(B − V) = 0.01 (determined using the Cardelli *et al.* 1989 extinction law). The stellar photospheric flux was then estimated by minimum $\chi^2$ fitting of the appropriate Kurucz stellar atmosphere model to optical and 2MASS (0.3 – 3 um) archival photometry (See URL http:// http://nsted.ipac.caltech.edu/). In the resulting excess spectrum, measurements in the 5.3 – 6.0 μm appear to be dominated by a strong upturn increasing towards shorter wavelengths (Figs. 1b, 4a-c) that is consistent with our team's 2-5 um NASA/IRTF SPeX measurements of scattered light from an icy dust excess, while the 34.8 – 35.3 μm range are very noisy and uncertain, so that for the thermal spectral decomposition described in this paper, we have used the more limited spectral range 6.0 – 34.7 μm as the measure of the goodness of fit. We have also masked out 2 very sharp, narrow spectral features, at 6.7 and 23.5 um, that are likely artifacts, but are mentioned here for completeness unless it is found in the future that they are instead narrow emission lines (e.g., the 6.7 μm feature may be an emission line of water). In any case, we do not model them for the results presented in this work. At the other extreme, from the upturn in the low frequency structure in the spectral baseline, there do appear to be potential broad, solid state emission features at ~12 and ~17 μm (Fig 4a,b), but at such low SNR that they do not affect the $\chi^2_\nu$ calculation used to find allowed models (i.e., within the 95% C.L.) significantly, and thus are not considered to be detected or detectable. Again, they are mentioned for completeness in case future work or analysis should show them to be of importance.

**2.3 Simple Interpretations.** The photosphere-removed flux is presented and compared to other significant mid-IR dust spectra (Fig. 1b, Fig. 2), and to the Deep Impact compositional model (Fig. 4). From our knowledge of the primary star and the amplitude of the excess spectrum and its run vs. wavelength, and our experience with other *Spitzer* mid-IR dusty disk spectra, we can discern a few things quickly: (1) the temperature of the warm dust is in the 300 - 400K range; (2) there are multiple fine emission features present; (3) the spectrum of the ~1 Gyr old *η Corvi* warm dust best matches the spectrum in our inventory that was measured for the ~10 My dust found around the young Herbig star HD100546, which we found to be in the process of disk clearing, comet aggregation and giant planet formation (Lisse *et al.* 2007a); (4) the spectrum is very rich in 5-8 μm emission, indicative of very primitive, carbon-rich, Ureilite meteorite-like dust (Sandford *et al.* 2010); and (5) the total mass of dust is relatively large, equivalent to the mass of the larger Centaurs or smaller icy moons and Kuiper Belt objects in our solar system; (6) the upturn in emissivity below 6 μm is strong, although, as noted, the SNR is low for these points at the edge of the SL order. The turn-up slope is much too steep to be due to thermal emission; it can, however, be well explained by high albedo, water-ice rich material with size distribution $dn/da \sim a^{-3.5}$ scattering light from an η Corvi primary with ~7000 $^{o}$K color temperature. The strong scattered light contribution is consistent with our team's detection of light scattering at 2-5 um from η Corvi by circumstellar material with Bond Albedo > 90 % using the SPeX instrument at the NASA/IRTF (Fig. 1b). It is also consistent with our analysis of the mid-IR spectra of cold, very ice-rich comets 17P/Holmes (Reach *et al.* 2010) and C/1995 O1 Hale-Bopp, and the disk of young Herbig star HD100546 (Lisse *et al.* 2007).

## 3. The Deep Impact Tempel 1 Dust Model

To further understand the information contained in the *Spitzer η Corvi* IRS spectra, we applied a more sophisticated spectral modeling analysis, based on the results of two recent solar system spacecraft experiments: Deep Impact and STARDUST. Here we summarize here the relevant portions of the Deep Impact experiment, its *Spitzer* IRS measurement, the Tempel 1 Dust Model created to interpret the measurements, and its predictions vs. the material returned by the STARDUST sampling experiment. Further details of the spectral analysis are described in the literature in the Supplementary Online material for Lisse *et al.* 2006, and in the main text of Lisse *et al.* 2007a.

**3.1 Deep Impact.** The recent Deep Impact hypervelocity experiment, one of the few direct man-made astrophysical experiments on record, produced a mix of materials - volatilized cometary material, silicaceous droplets, and excavated, largely unprocessed bulk material – after a small bolide of ~370 kg mass impacted the nucleus of comet 9P/Tempel 1 (hereafter Tempel 1) at 10.2 km $sec^{-1}$ on 2005 July 4, when the comet was 1.51 AU from the Sun (A'Hearn *et al.* 2005, Melosh 2006, Richardson *et al.* 2007). Due to the very low bulk modulus (< 10 kPa) and escape velocity (~ 1 m $sec^{-1}$) of the nucleus, the ejected material was heavily dominated, by 2-3 orders of magnitude in mass, by the excavation of bulk unprocessed material, as were the *Spitzer* emission spectra. This bulk material was excavated from the nucleus from depths as large as 30m and was largely unaltered, except for disruption of loosely-held macroscopic fractal particles into the individual sub-fractal, micron-sized components, as demonstrated by the presence of significant amounts of non-refractory water ice in the ejecta (Lisse *et al.* 2006; Sunshine *et al.* 2006). *Spitzer* IRS 5–35 μm spectra were taken within minutes of the impact, both before and after. The ejected material cooled from effects due to the impact within seconds to minutes, and separation of the ejecta emission spectrum from blackbody emission at LTE due to large, optically thick dust particles in the ambient coma was easily made. The resulting highly structured spectrum of the ejecta showed over 16 distinct spectral features at flux levels of a few Janskys (Lisse *et al.* 2006) that persisted for more than 20 hours after the impact.

**3.2 Thermal Emission.** The emission flux at wavelength λ from a collection of dust is given by

$$F_{\lambda,\mathrm{mod}} = \frac{1}{\Delta^2} \sum_i \int_0^\infty B_\lambda(T_i(a,r_*)) \cdot Emissivity_i(a,\lambda) \cdot \pi a^2 \frac{dn_i(r_*)}{da} \cdot da$$

where $T$ is the particle temperature for a particle of radius $a$ and composition $i$ at distance $r_*$ from the central star, Δ is the distance from *Spitzer* to the dust, $B_\lambda$ is the blackbody radiance at wavelength λ, Emissivity is the emission efficiency of a particle of radius a and composition $i$ at wavelength λ, *dn/da* is the differential particle size distribution (PSD) of the emitted dust, and the sum is over all species of material and over all particle sizes. Our spectral analysis consists of calculating the emission flux for a model collection of dust, and comparing the calculated flux to the observed flux. The emitted flux depends on the composition (location of spectral features), particle size (feature to continuum contrast), and the particle temperature (relative strength of short versus long wavelength features), and we discuss each of these effects below. N.B. - for ease of display, in order to uniformly emphasize the contribution of individual materials to the observed emission across the entire 5 to 35 μm wavelength range, and to facilitate comparison of the observations to laboratory emission spectra, the modeling results in this paper are presented in terms of emissivity spectra (e.g., a 390 K blackbody temperature dependence has been divided into the as-observed flux, best-fit model flux, and individual component fluxes in order to create the plotted emissivity spectra of Fig. 4).

**3.3 Composition.** To determine the mineral composition the observed IR emission is compared with the linear sum of laboratory thermal infrared emission spectra. As-measured emission spectra of randomly oriented, μm-sized powders were utilized to directly determine Emissivity(a,λ). The material spectra were selected by their reported presence in interplanetary dust particles, meteorites, in situ comet measurements, YSOs, and debris disks (Lisse *et al.* 2006). By building up a spectral library, we tested for the presence of over 80 different species in the T1 ejecta, in order to approach the problem in an unbiased fashion. We also expected, given the large number of important atomic species in astrophysical dust systems (H, C, O, Si, Mg, Fe, S, Ca, Al, Na, …) that the number of important species in the dust would be on the order of ~10.

The list of materials tested included multiple silicates in the olivine and pyroxene class (forsterite, fayalite, clino- and ortho-pyroxene, augite, anorthite, bronzite, diopside, and ferrosilite); phyllosilicates (such as saponite, serpentine, smectite, montmorillonite, and chlorite); sulfates (such as gypsum, ferrosulfate, and magnesium sulfate); oxides (including various aluminas, spinels, hibonite, magnetite, and hematite); Mg/Fe sulfides (including pyrrohtite, troilite, pyrite, and ningerite); carbonate minerals (including calcite, aragonite, dolomite, magnesite, and siderite); water-ice, clean and with carbon dioxide, carbon monoxide, methane, and ammonia clathrates; carbon dioxide ice; graphitic and amorphous carbon; and the neutral and ionized PAH emission models of Draine & Li (2007).

A model phase space search easily ruled out the presence of a vast majority of our library mineral species from the T1 ejecta. Only convincing evidence for the following as the majority species in the Tempel 1 ejecta was found (Lisse *et al.* 2006): crystalline silicates like forsterite, fayalite, ortho-enstatite, diopside, ferrosilite and amorphous silicates with olivine and pyroxene like composition; phyllosilicates similar to nonerite smectite, saponite, and talc; sulfides like ningerite and pyrrohtite; carbonates like magnesite and siderite; water gas and ice; amorphous carbon (and potentially native Fe:Ni; and ionized PAHs. This list of materials compares well by direct comparison to numerous in situ and sample return measurements (e.g., the Halley flybys and the STARDUST sample return; see Lisse *et al.* 2007a for a detailed list). These 7 classes of minerals (Ca/Fe/Mg-rich silicates, carbonates, phyllosilicates, water-ice, amorphous carbon, ionized PAHs, and Fe/Mg sulfides; 15 species in all), plus silicas, have now been successfully used to model thermal emission from dust emitted by 6 solar system comets, 4 extra-solar YSOs, and two mature exo-debris disks (Lisse *et al.* 2006, 2007a, 2007b, 2008, 2009). Specific sources for the emmisivity data used for the minerals identified in our best-fit model of the η Corvi circumstellar material include the Jena (http://www.astro.uni-jena.de/Laboratory/OCDB), MGS/TES (http://tes.asu.edu), and Glaccum (1999) spectral data libraries, as well as emission spectra supplied by Koike *et al.* (2000, 2002) and Chihara *et al.* (2002) for silicates; Kemper *et al.* (2002) for carbonates; Keller *et al. (2002)*, Kimura *et al.* (2005), and Nuth *et al.* (1985) for sulfides; Draine & Li (2007) for PAHs; Hanner (1984) and Edoh (1983) for amorphous carbon; and Koeberl (1988) for tektite silica.

**3.4 Particle Size Effects.** Particles of 0.1–100 μm are used in fitting the 5–35 μm data, with particle size effects on the emissivity assumed to vary as

$$1 - Emissivity(a, \lambda) = [1 - Emissivity(1\ um, \lambda)]^{(a/1\ um)}$$

The particle size distribution (PSD) is fit at log steps in radius, i.e., at [0.1, 0.2, 0.5, 1, 2, 5,…100] μm. Particles of the smallest sizes have emission spectra with very sharp features, and little continuum emission; particles of the largest sizes are optically thick, and emit only continuum emission. The *η Corvi* spectrum shows relatively strong, moderate contrast features, indicative of the presence of small (~1 μm) to medium (~20 μm) sized grains. A power law spectrum of particle sizes of moderate slope with dn/da ~ $a^{-3.5}$, dominated by both small & large grains, was found to be necessary to fit the *Spitzer* data.

**3.5 Particle Temperature.** We model the unresolved dust excess around *η Corvi* as a localized dust torus at a given astrocentric distance, with a unique value of temperature *T* for a particle of radius *a* and composition *i* (Eqn. 1). Dust particle temperature is determined at the same log steps in radius used to determine the PSD. The highest temperature for the smallest 0.1 μm particle of each species is free to vary, and is determined by the best-fit to the data; the largest, optically thick particles (100 μm) are set to the LTE temperature, and the temperature of intermediate sized particles is interpolated between these extremes by radiative energy balance, with a run of temperature vs. particle size similar to the calculations of Lien (1990).

The T1 ejecta were, by experimental design, all highly localized at 1.51 AU from the Sun. We use this fact to *empirically* determine the effective distance of the emitting material from *η Corvi*. To do this, a first-cut model temperature for the smallest and hottest (0.1 - 1 μm) particles for each material from our analysis is quickly determined from the strong constraints required to fit the high contrast spectral emission features, with a rough guess for the minimum temperatures of the largest and coldest dust. These first-cut maximum temperatures are compared to the temperatures found for the (0.5 - 2.0 μm) particles of the Tempel 1 ejecta (Lisse *et al.* 2006), using the relation

$$T_{dust} = T_{T1ejecta}(L_*/L_{solar})^{1/4}(1.51AU/r_*)^{1/2}$$

where $T_{dust}$ is the temperature of the smalles and hottest dust around *η Corvi*, $T_{T1\ ejecta}$ are as given in Lisse *et al.* 2006 (~ 340 K), $L_*$ = bolometric luminosity of *η Corvi*, = 4.9 $L_{Solar}$, and $r_*$ is the distance of the dust particle from *η Corvi*. Once we have $r_*$, we can determine $T_{LTE}$ for the 100 μm dust and interpolate using radiative balance to find the intermediate temperatures of the intermediate sizes; we then re-run our model to check our result, making minor changes in $T_{dust}$ and $T_{LTE}$ via iteration until we converge on an allowed solution fitting within the 95% Confidence Limit (C.L.) For this work the hottest particles were tested from 300 to 450 K for all species except for amorphous carbon (tested from 300 – 550K ), and water ice (tested from 150 - 220K). $T_{LTE}$ varied from 200 to 300 K.

**3.6 Model Summary.** Our method has limited input assumptions, uses physically plausible laboratory emission measures from randomly oriented powders rather than theoretically derived values from models of highly idealized dust, and simultaneously minimizes the number of adjustable parameters. The free parameters of the model are the relative abundance of each detected mineral species, the temperature of the smallest particle of each mineral species, and the value of the particle size distribution at each particle size (Table 2). Best-fits are found by a direct search through (composition, temperature, size distribution) phase space. To provide a statistical measure of the goodness-of-fit, we determine the 95% confidence limit given the number of degrees of freedom for a given reduced $\chi^2$ (see Lisse *et al.* 2008, 2009).

The total number of free parameters in the model used to fit the IRS *η Corvi* spectrum = 11 relative compositional abundances + 11 hottest particle temperatures + 10 particle size abundances = 32. [Since the results of our modeling show we could describe the dust within the 95% C.L. using 8 compositional abundances + 3 different refractory dust temperatures and 1 water ice temperature (Table 2) + 5 particle size parameters (3 blow-out + 1 overlap + 1 power law index) in actual practice we use only 17 adjustable values. With 1764 total data points, the difference between the 2 d.o.f. estimates is not critical for the $\chi^2_\nu$ 95% C.L. determination.] Based on these values, there is a 95% chance a model with reduced $\chi^2_\nu$ less than 1.06 is a good predictor of the Spitzer data. Our best-fit model successfully reproduces the entire IRS spectrum and yields a small $\chi^2_\nu$= 1.01; the allowed 2σ range for the derived parameters was found by determining the change from the best-fit value required to increase the chi-squared value to 1.06. The range of abundances from any model inside the 95% confidence limit is narrow, typically 10% from the best-fit value. There may be some bias for fitting the long wavelength data best, as there are 14 times as many SH/LH points at 9.9 – 34.3 μm than SL points at 5.2 – 9.9 μm, although the strongest and most highly structured (and thus highly constraining) features lie in the 6 – 12 μm range.

Our modeling has allowed us to get beyond the classical, well known olivine-pyroxene-amorphous carbon composition to the second-order, less emissive species like water, sulfides, PAHs, phyllosilicates, and carbonates. We are able to determine the overall amounts of the different major classes of dust-forming materials (olivines, pyroxenes, sulfides, water, etc.) and the bulk elemental abundances for the most abundant atoms in these materials (H, C, O, Si, Mg, Fe, S, Ca, Al). Applying our analysis, with a series of strong 'ground truth' checks of its validity, is highly diagnostic for interpreting mid-IR spectra of distant dusty systems like YSOs, debris disks, and PNs. It is important to note, however, that while our spectral modeling techniques can provide important new information, they are also complex. While designed to be as minimalist as possible, it took us over a year to analyze η Corvi properly and in depth – had Chen *et al.* 2006 used our more involved modeling methods, they would still be writing their 2006 paper today.

**3.7 Required Deviations From the 'Deep Impact Standard Compositional Model'.** As for our previous work on the unusual HD172555 spectrum (Lisse *et al.* 2009; Figures 3 & 4), silicas, indicative of the occurrence of high velocity impact, silicate transforming processes were required to fit the *η Corvi* spectrum. Even so, a linear sum of laboratory thermal infrared emission spectra did not entirely fit the *η Corvi* spectrum completely in the 7 – 9 μm region. I.e., no combination of emission from the Fe/Mg olivines, Ca/Fe/Mg pyroxenes, silicas, Fe/Mg sulfides, phyllosilicates, amorphous inorganic carbon, water ice/gas, PAHs, and carbonates commonly found in circumstellar dust, or from the other ~100 astrophysically relevant species in our spectral library was able to totally reproduce the observed spectrum. A number of new mineral species were studied and compared to the *η Corvi* spectrum, focusing on matching unusual emission features from 6.7 – 8.6 um, in the usual range for C-C, C=C, C-H, C-N, C-O, etc. vibrational modes. Lunar species such as anorthite and basalt, and all the allotropes of silica available in the literature were studied : quartz, cristoblite, tridymite, amorphous silica, protosilicate, obsidian, and various tektite compositions. Also examined were mineral sulfates (such as gypsum, ferrosulfate, and magnesium sulfate) and oxides (including various aluminas, spinels, hibonite, magnetite, and hematite), produced by aqueous alteration of primitive solar system materials, a distinct possibility for dust in an ~1 Gyr old system.

We found the high SNR of the *Spitzer* data in the 7 - 13 μm region to be highly constraining on the possible species present. As mentioned above, a good match to the *Spitzer* spectrum at the 95% confidence level (C.L.; $\chi^2_\nu < 1.06$) was found ***only*** in the singular case of a model spectrum which included very abundant silicates, silica, metal sulfides, and amorphous carbon, and these are the baseline "definitely detected" species we list in Table 2 and quote as the result from this paper. An even better fit which includes a mix of PAHs, fullerenes, HACs and nano-diamonds is possible, but one this is also very non-deterministic due to the very rich, complex, and varied chemistry of carbonaceous materials, and we consider modeling these beyond the scope of this paper. We do note, however, that the presence of copious HACs and nano-diamonds material, produced on Earth as the kinetic detonation product of carbonaceous materials in impact craters and industry (El

Goresy *et al.* 1999, 2003, 2004; Ross *et al.* 2010) is immediately suggestive of an impact origin for the circumstellar η Corvi dust, assuming a very carbon-rich starting material.

## 4. Modeling Results

**4.1 Dust Composition & Mineralogy.** Overall, η Corv's circumstellar dust appears to be rather crystalline (~50% in the silicates and silica). There are abundant silicates, metal sulfides, amorphous carbon, and water ice, typically found in cometary and other primitive icy bodies. New features at 6.8 - 7.3 um, 7.6 – 8um, and 8.5 μm and abundant silicas are found, that can be explained by moderate shock processing of a portion of the bulk material. The evidence for PAHs and water gas, also typically found in comets, is weak, but the SNR of the spectrum in the regions containing the characteristic emission features is low. On the other hand, there is no clear evidence for the presence of any carbonates, phyllosilicates, metal sulfates, or metal oxides due to aqueous alteration, arguing against the presence of warm, reactive water in the system for any great length of time, and it has too much C & S atomic content to have been heated and lost its volatiles in an equilibrium fashion over large periods of time.

**4.1.1 Silicates.** As determined from their 8-13 and 16 – 25 μm spectral signatures, *η Corvi*'s circumstellar silicates appear to be relatively primitive and Mg-rich, consisting of a combination of amorphous silicate, crystalline forsterite, and crystalline ortho-pyroxene and diopside, as found in comets. The dust is highly crystalline, > 70% by surface area, as seen for comets Hale-Bopp and Tempel 1 (Lisse *et al.* 2007b), and has been annealed vs. ISM material. On the other hand, the pyroxene:olivine ratio (Fig. 7) is too large for the silicaceous materials to have been modified by aqueous processing after incorporation into the parent body, as in dust derived from a carbon-rich C-type asteroid. We find a best-fit model with a spectrum ~45% due to silicates, of which about 2/3 are Mg-rich olivines and 1/3 are Fe-Mg-Ca-Al pyroxenes. Unlike our results on cometary systems (Lisse *et al.* 2007a, Reach *et al.* 2010, Sitko *et al.* 2011), amorphous olivine-like material is conspicuously lacking, and the amorphous pyroxene-like is present only in low abundance. Through low velocity impact processing, expected in KBOs (Farinella & Davis 1996, Stern 1996, Brown *et al.* 2010, Brown 2010), and in the larger impact that created the dust concentration at ~3 AU, it is likely that much of the primordial amorphous glassy silicates (typically in 1:1 ratio with crystalline materials), has been altered to form the observed crystalline silicates (Lisse *et al.* 2006) and silica.

**4.1.2 Silica.** Unlike comets and other primitive collections of dust, *η Corvi*'s dust shows evidence at 8 – 10 and 19 – 23 μm for a large fractional amount of silica dust, ~30% by relative surface area and ~50% by Si atom count, mostly in a high temperature, low pressure quick-quenched glassy tektite-like phase. Dust in the ~12 Myr, A5V HD172555 system showed an even larger preponderance (> 50% by surface area and 70% by Si atom count) of silica, argued by Lisse *et al.* (2009) to be formed during a giant hypervelocity ($v_{interaction}$ > 10 km/sec) impact between two large planetesimals or planets. Silica is the kinetically favored species produced by intense, rapid heating and melting/vaporization of silicate-rich bodies followed by quick quenching - silica-rich materials are found in terrestrial lava flows and in the tektites and glasses found at the hypervelocity ($v_{impact}$ > 10 km/sec) impact craters on the Earth and Moon (Warren 2008). The key method of detecting the silicas is that they are the only source of strong, broad IR emission features at 8 - 10 μm ***and*** 20 - 22 μm. The tektite standard used for comparison here, a bediasite from Texas (Koeberl 1988), represents one of the 2 major tektite classes found on Earth. The Tektite materials also contain minority fractions of MgO, FeO, CaO, $Al_2O3$, $Na_2O$, and $K_2O$; these minor fractions do not affect the spectral signature overly compared to changes seen due to condensing the material into the different polymorphs of silica, which can easily shift the emission peaks by 0.5 μm while changing their shape. Because of this effect, we are able to determine that crystalline silica phases such as cristobalite and tridymite are not present, consistent with the lack of a significant 12.6 μm feature found for T Tauri systems rich in crystalline silica (Sargent et al. 2006; W. J. Forrest 2008, private communication). On the other hand, it is possible that a little bit of crystalline quartz is present as a minor species, but only just at the 95 % C.L. (Table 2).

Unlike HD172555, the *η Corvi* dust does not show the majority of dust to be in the silica form; there is about as much olivine and pyroxene in the mix. Models of impacts argue that melting and vaporization of 2 bodies will be incomplete and most concentrated near the point of contact for relative impact velocities of 1 to 10 km/sec, with the efficiency of melting and conversion to silica rising rapidly at higher velocities, and becoming negligible below 1 km/sec (Grieve & Cintala 1992, Cintala & Grieve 1998), the relative velocities expected for impact processing in the Kuiper Belt (Farinella & Davis 1996, Stern 1996); ≤ 5 km/sec are the impact velocities found in our asteroid belt. Finding approximately one-half of all η Corvi's dust in silica form argues for higher collisional interaction velocities, on the order of 5 to 10 km/sec, marginally possible in an η Corvi asteroid belt at ~3 AU from the primary, and definitely possible between a KBO and a body at ~3 AU (See §5.5). As silica is not found in dominant abundance in interplanetary dust particles in our own solar system, nor is obvious in the large majority of mature debris disk spectra (e.g., HD69830, Lisse *et al.* 2007b; ε Eridani, Backman *et al.* 2009) this would appear to be good evidence that the η Corvi system currently has an unusually dynamically excited debris disk. A high velocity impact source for the η Corvi dust is also consistent with the detection of shock-produced nano-diamonds from amorphous carbon and fullerenes from PAHs (see below).

**4.1.3 Phyllosilicates.** No phyllosilicate spectral signatures at ~10 μm, usually associated with aqueous alteration of silicate species, are detected in the η Corvi spectrum, arguing for a very primitive parent body source which never came in contact with a source of warm, reactive water. This is rather surprising given the large amount of water ice and gas detected in our spectrum, and our conclusion of a medium sized KBO or large sized Centaur parent body for the η Corvi dust. It is consistent with the non-detection of carbonates, sulfates, and metal oxides, also products of aqueous alteration. From a meteoritic perspective, a possible answer is that we are observing moderately shock heated carbonaceous chondrite material from a C-type asteroid with preferential destruction of phyllosilicates and hydrous sulfates in favor of olivines, pyroxenes, and sulfides (Tomioka *et al.* 2007, Morlok *et al.* 2008, 2010), but such material contains too little carbon, no silica, and abundant magnetite compared to what we find in the η Corvi dust. We are driven to conclude instead that there was little to no free, warm water for any substantial amount of time in the parent body or bodies of the η Corvi dust, and that the body was assembled at very low temperatures where water was frozen out and unreactive, far from the η Corvi primary. This is consistent with a Kuiper Belt origin for the parent body of the dust, & inconsistent with a close-in asteroidal source (see Fig 2).

**4.1.4 Metal Sulfides.** We detect a large abundance of metal sulfides due to strong mission at 25 – 35 μm in the *Spitzer* η Corvi IRS excess spectrum, similar to what is found in cometary and HD100546 dust and the STARDUST sample return. As for cometary systems, the major source of iron is in the Fe-sulfides (pyrrohtite, ningerite, and pentdantalite), and there is some minority iron in the iron-bearing olivines. There is not obvious signature of the minority iron bearing carbonate species siderite ($FeCO_3$). The main sulfur reservoir in the *η Corvi* dust is the metal sulfides. The early solar system was a strongly reducing environment, as expected for a system with >1000 H atoms for every O atom, and evidenced by the high abundance of Fe-sulfide (sulfur was the major oxidant available in the early solar system, condensing out from S vapor at ~600K and attacking any free Fe), but not Fe-sulphate or Fe-oxide species in primitive cometary dust (Zolensky *et al.* 2006, Lisse *et al.* 2007a). We do not find evidence for Fe-oxides, and only very faint (at the few % C.L.) evidence for Fe-or Mg-sulfates, as expected in warmed icy moon-like bodies with subsurface water-rich oceans (Postberg *et al.* 2009). Sulfates and oxides are the normal end product of aqueous oxidative attack on sulfide materials on the terrestrial planets, but the oxidative attack on native Fe stopped at the sulfide level in the PSN.

The lack of sulfates and oxides in the *η Corvi* dust, the end-state oxidation products of metal sulfides found on the surface of terrestrial planets with copious available water (e.g. Mars or the Earth), along with the lack of obvious phyllosilicates and carbonates, argues against substantial aqueous alteration in an oxidizing environment of the parent body. We can use the same non-detections to rule out a plate-tectonic, organics rich, aqueously altered terrestrial planet surface as the source of the observed dust, as reasonably suggested by Fujiwara (private comm., 2009). [On Gyr timescales, silicates are converted into crystalline silica, phyllosilicates, oxides, sulfates, and carbonates on the modern day Earth via plate tectonic subduction mixing hot rock with available reactive $CO_2$ and water, while Mars shows extensive evidence of Gyrs of aqueous alteration in its layers of oxides, sulfates and phyllosilicates over basaltic olivine & pyroxene (B. Thompson 2010, private commun.)]

**4.1.5 Amorphous Carbon, PAHs, and Other Carbon Bearing Species.** At 5 - 8 μm, amorphous carbon emission is present in our best fit model at very high relative abundances, akin to those found for comets and HD100546, the other carbon-rich systems we have studied to date (Lisse *et al.* 2006, 2007a; Figure 5). As there is currently some uncertainty in the removal of the stellar photospheric contribution which is critical to the detection of amorphous carbon, we quote here the range for the abundance of amorphous carbon in the debris disk by allowing the photospheric model to vary between the extremes of the allowed 2σ limit of normalization. We conclude that the amount of amorphous carbon in the circumstellar dust to be between 0.14 and 0.20 (by relative surface area), implying a significant amount, 2.9 to 4.5 moles (relative), or between 43 and 54% by mole fraction of the total circumstellar material.

Emission lines from carbon-bearing PAHs were detected in the 6 – 9 μm region of the η Corvi spectrum. Compared to HD100546, they are found at modest relative amplitudes, but compared to comets, at typical levels. As described in Lisse *et al.* (2007a), this may be due to a relative paucity of IR-fluorescence exciting UV photons emitted by the η Corvi F2V star versus the A0V primary of HD100546. There are also new features at 6.8 - 7.3 um, a potential feature at 7.6 – 8 um, and a very strong 8.6 μm peak that cannot be accounted for by emission from silicas or PAHs. (While PAH emission typically show strong features at 7.7 and 8.6 μm, these emission are coupled, and the line ratio for ionized PAHs such that the 7.7 μm emission is much stronger than the 8.6 μm emission.) For η Corvi we find much the reverse, a very strong contribution at 8.4 to 8.6 μm. Sulfates, which have a very strong emission feature at 8.6 um, have other emission features in the 9 – 20 μm range that are not seen and thus cannot be contributing significantly to the observed flux. The 6.8 - 7.3 μm feature is in the 6.5 – 7.2 μm region typical of carbonate emission, but is narrower and weaker than the typical carbonate features seen by us in cometary spectra, and the corresponding 11.2 and 14 μm features are not found.

Instead, we note that this region is rich in vibrational emission features from organic carbon species potentially seen in our η Corvi emissivity spectrum, including hydrogenated amorphous carbon (HACs) at 6.8 and 7.25 um, surface moieties on and impurity modes in nanodiamonds at 7 – 7.4 μm and 8.5 – 9.0 μm (Fig 4d), fullerene stretches at 7.1, 8.5, 12.7, 14.8, 15.7, 17.2, 18.8 μm (Fig 4e), and alkane out-of -plane bending modes at 6.8 and 7.4 μm. All of these $sp^3$ and $sp^2$ species are consistent with a large relative abundance of carbon-rich species created by shock heating of primitive mix of amorphous carbon, nano-graphite, and simple hydrocarbon species (Stroud *et al.* 2011) or alteration of organic ices found on outer solar system bodies (e.g., tholins, Emery 2010 priv. comm., Fig 4f). Support for this finding is seen in the rare solar system Ureilite meteorites (e.g. Almahata Sitta), which show a very similar mid-IR spectrum (Fig 4) and very high abundances of carbon and organics, especially in graphitic and nano-diamond phases and complex organics (Sanford *et al.* 2010, Zolensky *et al.* 2010), purportedly due to shock and heating of an unknown but very primitive C-rich parent body in the very early solar system. The Ureilites consist predominantly of ferromagnesian olivine and pyroxene, as well as Ca-rich pyroxene (like diopside) in various proportions, in addition to minor elemental carbon - as found for cometary bodies (Lisse *et al.* 2007a, Hanner & Zolensky 2009). Note, however, that other, less carbon rich meteorite samples, like the carbonaceous chondrites Allende, Murchison, Ornans, and Virgano (Fig 4), do not match the *Spitzer* η Corvi spectrum well.

Determining exactly which of these carbon species is present, though, can be challenging using remote sensing, due to the richness of carbon organic chemistry; for example, a perusal of the literature on nano-diamond mid-IR absorptance spectra will return as many different spectra as papers. (The main reason for this is that the absorptance behavior of nano-diamond is dominated by minority impurities at the ppm level and surface functional groups highly dependent on the chemical methods used to isolate the nanodiamonds.) For the purposes of this first results paper, because of the current uncertainty in the source of these emissions, we have chosen to leave these narrow emission features unfit. ***Regardless, the critical point is that emission in this region is ONLY found in carbon rich, primitive dust dominated Herbig Ae disk and comet dust spectra, and not in any of the rocky or asteroidal parent body debris spectra dominated by Si-O bonds that we have studied to date (Fig. 2). In fact, the emission lines in this region are stronger than those found in comets and most like those found in the most primitive of observed dust collections in young Herbig Ae/Be disks. Reflectance spectra of solar system KBO surfaces show that this material is also organics rich (Barucci et al. 2005, Dalla Ore et al. 2009).*** A large Centaur or Kuiper Belt object (r > 100 km), could easily retain abundant primitive carbonaceous material by maintaining a combination of low formation temperature and relatively large escape velocity. (e.g., the escape velocity for Enceladus, the icy moon of Saturn with roughly cometary composition and r = 235 km is ~220 m/sec, as opposed to the ~1 m/sec escape velocity for the 3 km radius comet 9P/Tempel 1).

**4.1.6 Water Ice and Gas.** Water ice in considerable amounts is found in our best-fit model. The clear presence of water ice at 11 to 15 μm is somewhat surprising at first look, given that the smallest dust particles appear to be as hot as 350K, and the average large dust particle temperature (~LTE) is 250 K, somewhat above the ~200K temperature at which water ice freely sublimates at 0 torr. Lisse et al. (2006) found the same dichotomy in refractory vs icy dust temperatures for the Tempel 1 ejecta produced by the Deep Impact experiment. Cooling by evaporative sublimation can stabilize the ice if it is isolated from the other hot dust, is "clean", and/or is present in large ( > 1 mm) chunks (Lien 1990).

One of the first considerations is whether this water ice is coincidentally in the beam with the warm dust at 2 AU, and arises from the same ~150 AU source as the sub-mm dust detected by Wyatt et al (2005). Herschel has detected large cold dust at ~100 – 150 AU (Matthews *et al.* 2010, Fig 1c). Water ice at this location would have temperatures ~35K, however, and would be undetectable in the 5 - 35 μm wavelength range of the IRS (hard on the Wien law side of a 35 K blackbody, Eqn. 1) unless present at extraordinary quantities – a situation which is ruled out by measured SEDs (Figure 1; a similar analysis was used to distinguish water ice at different temperatures in the HD113766 system (Lisse *et al.* 2008)). Further, HST has not detected any signature of fine water ice in the system's Kuiper Belt (Clampin & Wisniewski 2011, priv. commun.), so any fine water ice detectable in the infrared must be located within an HST PSF, or inside ~ 10 AU of the primary.

Instead, we must surmise that the water ice component in our best fit model, at temperatures of 170 to 210 K, is located in the inner reaches of the η Corvi system, likely mixed in with the warm dust at ~3 AU, with its $T_{LTE}$ ~250K. While surprising, this explains the strong upturn we find in the *Spitzer* IRS emissivity shortward of 6 μm that cannot be fit by a Kurucz stellar emission model (Fig 4a), as well as the strong blue scattered light excess detected at 2 to 5 um by our team using R~1000 spectroscopy from the NASA/IRTF SPeX instrument (Fig 1b). This excess is well fit by a water ice scattering model and demonstrates possible absorption features at 3 and 5 um. These findings all appear to be good evidence (after careful examination and maximal possible stellar photospheric removal) for abundant high albedo ($p_v$ > 50%) water ice particles close into the star (i.e., inside the HST limit of ~10 AU; Bryden *et al.*'s 2009 location of all warm dust within a few AU of the primary; and Smith *et al.'s* 2009 limit of all warm dust within 3.5 AU of the star).

***Creation of stable water ice particulates by an impact mechanism is plausible***; the presence and survival of ejected water ice from a 10 km/sec impact on an ice-rich body was demonstrated by the Deep Impact experiment at 10.2 km/sec relative onto comet 9P/Tempel 1 in 2005 (Sunshine et al. 2006, Gicquel *et al.* 2011), and creation of actively sublimating large water ice particles from an icy primitive body was directly demonstrated during the recent EPOXI flyby of comet 103P/Hartley 2 (A'Hearn *et al.* 2011). ***Survival of water ice over many years is another story*** - Bockelee-Morvan *et al.* (2001) estimate the lifetime for a water ice grain to be given by $\tau_{sublimation}$[sec] = 2.4 x $10^4$ sec * a(cm) at $r_h$ = 1.08 AU from the 1 $L_\odot$ Sun, assuming $\rho$ = 0.5 g $cm^{-3}$ and "dirty" ice with an ice-to-dust mass ratio κ = 1. This implies that the fine 0.1 μm to 1 cm icy dust released from a normal comet via outgassing or impact cratering (Lisse *et al.* 1998, 2004, 2006, 2007) would dissipate on the order of days or less due to evaporation via stellar radiation driven sublimation; it would take a huge piece, ~10 km radius, of ice (extrapolating hugely across 6 orders of magnitude in behavior) just to survive on the order of 1000 yrs, the age of the observed circumstellar dust that we estimate below. We have some guidance for long-term water ice stability from observations of evaporating and fragmenting icy comets (a = 0.3 to 50 km) in our solar system; we know that the icy cometary bodies lose material upon close passage through the THZ of the Sun, from fine μm sized dust up to meter-sized boulders, yet survive for thousands to millions of years (Lisse *et al.* 2002) and are covered with dark, low albedo surface material ([i.e., like the mantle on the surface of icy comets allowing the 300 – 400 K surface temperatures observed on comet Halley (11 km radius, lifetime > 2500 yrs, Krasnopolsky *et al.* 1987); Borrelly (2.4 m radius, lifetime > 300 yrs, Solderblom *et al.* 2004); and Tempel 1 (3 km radius, lifetime > 150 yrs, Groussin *et al.* 2007).

There is possible evidence for a water gas detection in our model, at the ~5% relative abundance level, suggestion gas production by heating and sublimation of the water ice present. However, the water gas was only marginally detected in our *Spitzer* measurement, and we do not consider our analysis definitive on this subject, in that the IRS data become very noisy in the 5 - 7 μm region diagnostic of water gas as well as carbonate, PAHs, amorphous carbon, and other carbonaceous materials. The presence of a weak water gas feature, if real, would be reassuring, as it suggests some active production of water ice and sink of it into water gas, followed by its quick dissociation (in ~20 hrs at ~3 AU) by the F2V primary's UV emission into O and H atoms followed by ionization, pickup, and outsweep by the primary's stellar wind. Further, deeper searches for $H_2O$ gas, and OH and OI daughter products are warranted. On the other hand, we can clearly rule out the presence of large amounts of free water gas in the η Corvi system.

Overall, the presence of copious amounts of circumstellar water ice argues that something must be storing these grains and releasing them continuously, as would large objects being continually ground down, or something must be working to extend the lifetimes of the grains. The former scenario is inconsistent with the lack of a strong 6 μm water gas emission feature in the Spitzer spectrum, and also implies many orders of magnitude more icy dust mass than we detect today, so it is probably more likely that the water ice grains are very long lived. Very small grains, << 0.4 um, can decouple from the local radiation field and last longer; however, these grains would not emit well at wavelengths of 2 - 35 μm by the same effect, and thus would have to be in huge abundance.

Another possibility is that the observed water ice is very pure, without refractory dust inclusions or mixtures, so that the absorption of incident starlight is very low and heating by insolation is miniscule. Lien 1990 calculated that this effect can extend the lifetime of water ice particles by many orders of magnitude, finding lifetimes of $10^3$ to $10^7$ years for pure grains of radius 10 and 100 um, respectively. (The lifetime of a grain is a strong function of its purity, and it takes as little as 5% by mass of darkish refractory material to maximize the "dirtiness " of a water ice grain and shorten its lifetime to its asymptotic minimum.) These pure water ice grain lifetimes are consistent with the detection of 0.1 – 100 μm particles by Spitzer that we report here and the ~$10^3$ yr we estimate have elapsed since the formative collision (see Sec. 4.5). The presence of pure water ice grains in Comet 9P/Tempel 1 and η Corvi argues strongly for a condensation mechanism for water from the gas phase that is independent of the solid refractory phase in an icy primitive body, and also greatly extends the amount of time icy grains have available to be incorporated into the body during its formation.

**4.2 Elemental Abundances.** The major refractory species, with the exception of S and Al, are all very depleted vs. solar (Fig. 5), a pattern similar to those found for the young debris disk system HD172555, with silica dominated debris formed as a result of a giant hypervelocity impact, and the Herbig disk system, HD100546. The elemental abundance pattern is in distinct contrast to the near-solar values for cometary dust in our remote sensing assays (Lisse *et al.* 2006, 2007, Reach *et al.* 2010, Sitko *et al.* 2011) and from the STARDUST sample return (Brownlee *et al.* 2006, Hanner & Zolensky 2009). The abundance pattern similarity to the hyper-velocity created dust in the HD172555 system argues for significant collisional processing of the *η* Corvi warm dust, with concomitant preferential removal of Fe, S, and Ca, and addition of Al. A similar pattern of elemental processing is seen between the primitive upper mantle rocks of the Earth and melt-altered crustal material (e.g., Adam, Baker, & Wiedenbeck 2002, and references therein).

**4.3 Temperature of the Dust and Dust Location with Respect to the Primary.** In previous work, long-wave infrared observations were used to reveal an extended disk around *η* Corvi of cold dust grains, located in a region intriguingly similar to our Sun's extended "Kuiper Belt," a zone of small bodies left over from the formation of our Solar System and the reservoir of short-period comets. Similar disks are seen around β Pictoris (Okamato *et al.* 2004, Chen *et al.* 2007), Fomahault (Holland *et al.* 2003, Kalas *et al.* 2009), and ε Eridiani (Greaves *et al.* 1989, Backman *et al.* 2009). Wyatt *et al.* 2005 found η Corvi to possess a cold dust component with $T_{dust} = 35 \pm 5$ $^{o}$K, as inferred from SED fits to IRAS 12 - 100 μm and JCMT SCUBA 450 and 850 μm photometry. Their images of the 850 μm submm dust ring locate it at 150 ± 20 AU, where $T_{LTE}$ = 35 K. The inner 100 AU appears clear of sub-mm dust emission (c.f. Bryden *et al.* 2009; Fig. 1).

The IRAS IR excess for *η* Corvi argues for a warm dust component of 370 ± 60K at ~2 AU from the primary. A second component is detected in IRAS photometry; self-consistent modeling of the IRAS and submillimeter photometry suggest that this population possesses a grain temperature $T_{gr}$ = 370 ± 60 K (Wyatt *et al.* 2005). Detailed fits to the *Spitzer* IRS 5.5 - 35 μm spectra by Chen *et al.* (2006) suggested that this emission is produced by amorphous olivine, crystalline olivine (forsterite) and crystalline pyroxene (ortho-enstatite) features with $T_{dust}$ = 345 K and a crystalline silicate fraction of 31%, and black body grains with a temperature $T_{dust}$ = 120 K.

In this work we find the hottest small silicate grains, our best "thermometers" from the Deep Impact experiment (§3.5) to be at 350 ± 20 K (2σ), implying a dust location of 3.0 ± 0.3 AU (2σ) from the η Corvi F2 primary if we assume $L_* = 4.9\ L_{Solar}$. ***This is the equivalent of being ~ 1.3 AU from the Sun in our solar system, a bit farther out than the Earth but still within the THZ and the realm of liquid surface water on any extant planetesimal.*** By comparison, *Herschel* observations reported by Matthews *et al.* (2010) place the η Corvi cold dust at 100 – 200 AU and the warm dust at 1.4AU (but assumes all dust is at 370 K and at LTE temperatures, ignoring the possibility of superheated dust. A similar "moving out" of warm dust detected in the HD69830 system was found by our group – initial reports of dust at ~0.6 AU from the primary were updated to dust at ~1 AU from the primary, after allowing for super-LTE heating of small dust grains) while Smith, Wyatt, & Dent (2008) and Smith, Wyatt, and Haniff (2009) using VLT/MIDI mid-IR interferometry place the warm dust at 0.16 – 3.5 AU (Fig 1; we note that our locating the warm dust at 3.0 ± 0.3 AU from this work is consistent with the outer possible extent found by these groups; if correct, this implies that future measurements with larger telescopes or deeper integrations may be able to resolve the η Corvi warm dust population). We are confident in our estimates of the location of the η Corvi warm dust, as we placed the dominant warm dust in HD100546 at ~13 AU from the primary (Lisse *et al.* 2007a), similar to the size of the inner gap found by Grady *et al.* (2005), and since comparison of recent observations of the warm dust location in HD 69830 and HD113766A using MIDI at the VLT (Smith *et al.* ref. 2009, 2011) show them to be consistent with the estimated locations derived by our group using the same method (Lisse *et al.* 2007b, 2008).

A third dust reservoir, associated with the 120K blackbody dust at ~12 AU, reported in Chen *et al.* 2006, has not been reproduced here. From our work, we do not consider the case for this dust to be conclusive, and finding a different modeling outcome to be reasonable given that (1) the re-extracted and re-calibrated IRS mid-IR spectrum for η Corvi has changed substantially (see §2.2); (2) here we have improved over the first-order spectral models of Chen *et al.* (2006) with a focused in-depth model of each spectral feature, applying 5 more years of study of primitive body compositional mineralogy derived from analysis of the STARDUST sample return, remote sensing measurements made during the Deep Impact experiment, and thermal emission and transmission measurements of the emissivity of laboratory analogue powders; and (3) Bryden *et al.* (2009) find no evidence from interferometric imaging for an appreciable warm dust population outside a few AU in the η Corvi system, while Smith *et al.* (2009) localize the dust inside the Kuiper Belt to within 3.5 AU of the primary.

The model of Wyatt *et al.* (2010), invoking an eccentric ring of dust with pericentre at 0.75 AU ($T_{LTE}$ = 500 K) and apocentre at 150 AU does not fit the warm dust temperatures and *Spitzer* mid-IR spectrum we examine here, unless the dust is spread out along the orbit so as to appear to have a weighted color temperature in the 250 – 350 K range of our best fit model.

**4.4 Particle Size Distribution and Mass of Dust.** From the modeling performed in this work, we find that the best-fit model particle size distribution is $\mathbf{dn/da \sim a^{-3.5 \pm 0.01}}$ above 1 μm, as expected for dust in collisional equilibrium, and falls off faster than this below ~1 μm, as expected for dust particles smaller than the few micron blowout cutoff of the system (Figure 6). (Note that our empirical blowout size is slightly smaller than Wyatt *et al.*'s 1999 blackbody value of 4 μm, but here we utilize the opacities for real dust materials.) ***Both of these findings argue for dust that has had time to come to a steady state equilibrium by balancing radiation forces with collisional grinding of impact fragments. i.e. dust that is more than a few 10's of orbits old (~$10^3$ yrs at 5.8 AU around an $1.4M_\odot$ star).***

Integrating this PSD across particle sizes from 0.1 to 100 μm, and normalizing to the total absolute fluxes measured for the circumstellar excess around η Corvi, we find a total minimum warm dust mass of > 8.6 x $10^{18}$ kg. This is an interesting amount of mass – larger by far than any comet ever observed in the solar system, but quite close to the mass estimated for a small to medium sized icy moon of the giant planets, a larger Centaur body, or a small to medium sized KBO (Table 3). It is also > 250 times the mass of dust in our current zodiacal cloud, assuming the cloud is the equivalent of one 15 km radius asteroid of 2.5 g $cm^{-3}$ density. We note that there is nothing special about a largest particle size of 100 μm, other than it is the largest particle easily measured by *Spitzer*; it is highly plausible that fragments upwards of 100 m exist, which would imply 1000 times more total mass, or ~9 x $10^{21}$ kg of warm dust, equivalent to the mass found in the largest KBOs of the solar system.

These results can be usefully compared to estimates derived for the extended cold dust component of the η Corvi disk. This component is well fit by emission at $T$ = 35 ± 5 K, $b$ = 0.5 and $l_0$ = 20 μm. The fractional luminosity of this component is $f = L_{IR}/L_* = 3 \times 10^{-5}$ and the inferred dust mass assuming an opacity of $k_{850\mu m}$ = 0.17 $m^2$ $kg^{-1}$ (Wyatt, Dent & Greaves 2003 and references therein) is 0.04 $M_{Earth}$ for particles of 0.1 – 1000 μm size. This is roughly 2.4 x $10^{23}$ kg, or a factor of 3 x $10^4$ more mass than we find here for the η Corvi warm dust; it is about the amount we would expect, though, from a simple scaling of the two dust SEDs for η Corvi that we see in Figure 1 - each of roughly the same amplitude in flux but varying by a factor of 10 in color temperature. (Assuming Flux ~ $\sigma T^3$, we would expect there to be about $10^3$ times more emitting surface area and 3 x $10^4$ times more mass in the cold dust reservoir than in the warm). It is compelling to note that it would take only ~1/30000 of the mass of the observed η Corvi Kuiper Belt to produce the warm dust detected – dynamical stirring of the system's Kuiper Belt should perturb icy planetesimals onto many different orbits, and it would take only a very small fraction, or even just one, of the scattered planetesimals to provide the observed warm dust.

Using a simple model (with collisional lifetimes of planetesimals computed using a model for the outcome of different sized collisions, and assuming a size distribution appropriate for material in a collisional cascade scaled to the surface area of dust in that distribution through a fit to the SED of the disk emission), Wyatt *et al.* 2005 inferred from the age of 1 Gyr that the collisional cascade in this system starts with planetesimals a few km in size, implying an total mass of $20M_{Earth}$ in planetesimals in the ring at 150 AU. This is similar to their inferred mass for the collisional cascade of the disk around the star Fomalhaut at age ~200 Myr (Wyatt & Dent 2002).

In a related study using a more detailed SED model using realistic grain properties, Sheret *et al.* 2004 found that the cold dust SED could be fit with a dust ring at 150 AU and a collisional cascade dust size distribution, but one with an additional imposed cut-off for grains smaller than $a_{min}$ = 30 μm rather than at the $a_{min}$ = 4 μm expected from a simple blackbody radiation pressure blow-out calculation (Wyatt *et al.* 1999). Grains in the size range of 4-30 μm must be absent because they are too hot (>> 35 K) at this distance from the star to explain the shape of the SED, which would have stronger emission at 25 and 60 μm if these grains were present. Possible reasons for the absence of small grains, also seen in the spectrum of *ε* Eridani, were discussed in Sheret *et al.* (2004). These include the possibility that the 4-30 μm grains in the outer disk of *η Corvi* are destroyed in collisions with < 4 μm grains which are in the process of being blown out of the inner regions by radiation pressure (Krivov *et al.* 2000, Grigorieva *et al.* 2007, Czechowski & Mann 2007). We also note that a dearth of small grains is expected in the model in which clumpy structure is formed by planet migration (Wyatt 2003, 2006; Figure 3). ***Both of these results are consistent with the PSD we find here for the warm dust in η Corvi (Figure 6).***

**4.5 Timescale for the η Corvi Dust.** The best-fit PSD for the η Corvi dust has structure in it due to the interplay of stellar radiation, stellar gravity, and dust-dust collisions. This structure can be used to estimate the dust's minimum age. E.g., the timescale for dust to be removed via radiation pressure and P-R drag (the so-called "blow-out time") at 3 AU is essentially the time it takes for dust to complete an orbit – i.e., the dynamical time. For the warm dust, this is on the order of few years at 3 AU around an F2V star, and warm dust less than a few μm in radius is quickly removed from the system within 3 to 6 yrs after its formation – so that any sub-micron dust we see today has been re-supplied via collisional grinding of the initial impact fragments (see below). We can also use the radiative timescale line of reasoning to "date" the cold dust in the Kuiper belt. Clampin & Wisniewski (2011, priv. commun.) report that extended emission around η Corvi was not seen in the optical by HST, implying that the extant Kuiper Belt dust detected by SCUBA and Herschel (Fig. 1c) must be large and also old, in that young fresh dust would have a detectable small particle component. Thus the Kuiper Belt collisions are not very recent, i.e. they have to have occurred at least more than a blowout dynamical time ago, > 1000 yrs at 150 AU.

The timescale for dust to reach the collisional steady state found for η Corvi (whereby large particles grind down by collisions to create small ones) is their collisional lifetime. The time to reach a collisional steady-state depends on the local optical depth but also on the maximum particle size of the population. Collisional steady-state is reached first at the smallest sizes (because they contain most of the geometrical cross section) and then works its way up the size distribution (Wyatt *et al.* 2011**)**. If we make the fiducial assumption that all the $3 \times 10^{19}$ kg of dust is distributed in an annulus 1 AU wide centered around 3 AU, and distributed from 1 μm to 1 mm particles following the best-fit $a^{-3.5}$ power law, we find a corresponding vertical optical depth of $\sigma \sim 0.0015$ and an approximate collisional timescale of $\tau_{orb}/(12\sigma) \sim 60\tau_{orb}$. Conservatively then, ~100 orbital timescales or ~ 1000 yrs is needed to reach a collisional steady state for the detected micron size grains surviving blowout. This timescale is consistent with the radiative lifetime estimate for the cold Kuiper Belt dust formation derived above, implying that whatever formed the two dust belts in η Corvi happened over 1000 years before the *Spitzer* observations of the system.

## 5. Discussion

**5.1 Nature of the Warm Dust.** As stated in §4, the best-fit model for the η Corvi warm dust contains materials typical of primitive cometary dust, i.e., Fe/Mg olivines, Ca/Fe/Mg pyroxenes, silicas, Fe/Mg sulfides, amorphous inorganic carbon, water ice, and PAHs. Abundant silica and features suggesting the presence of carbon-rich material similar to nano-diamonds, fullerenes, and HACs, produced in collisions with V > 5 km/sec, are also present. The dust appears to be very crystalline, as seen for many comets, and especially crystalline in the silicates. The temperature of the hottest non-carbon dust particles, < 1 μm in size, is ~350 K, placing the dust at 3.0 ± 0.3 AU in the η Corvi system, or at the equivalent of ~1.3 AU in our solar system, near the outer edge of the solar system's THZ. The size distribution for the dust trends as **dn/da ~ $a^{-3.5}$** above 1 μm, as expected for dust in collisional equilibrium, and falls off faster than this below 1 μm, as expected for dust smaller than the blowout cutoff of the system, implying dust in steady state equilibrium and older than 1 Kyr. The dust has been in its present location for over 1000 yrs. The total mass of dust is at least 9 x $10^{18}$ kg, equivalent to a 126 km radius, 1.0 g $cm^{-3}$ non-porous icy body (medium to large Centaur, medium sized KBO) or 172 km radius, 0.4 g $cm^{-3}$ porous icy (gigantic & unknown in the solar system) comet-like body.

Important constraints on the nature of the η Corvi warm dust can also be obtained using the basic debris disk model of Wyatt *et al.* (2007). When applied to the η Corvi system, the results show that there is much too much warm circumstellar dust for a ~1 Gyr old system, and we can rule out a self-stirred or massive disk. This result is in qualitative agreement with the statistical argument presented in §1, where we noted that the η Corvi excess (over the photosphere) flux is ~1000x brighter at 24 um than any other known system at ~1 Gyr. We must then invoke a stochastic collision scenario to explain the observed warm dust. Wyatt *et al.'s* 2002, 2005, 2007 studies of the η Corvi's cold dust came to a similar conclusion, with the exception that it is not one, but many collisions that are occurring, and that there is likely something dynamically perturbing the system's Kuiper Belt. From the lack of small cold Kuiper Belt dust Clampin & Wisniewski (2011, priv. commun.), we can infer that the dynamical time for collisions in the active and stirred up Kuiper Belt is longer than 1000 yrs (Sec 4.5), consistent with the dynamical age of ~ $10^3$ yrs for the warm dust at ~3 AU inferred from its PSD.

**5.2 Comparison to Spectral Studies of Other Bright Dusty Disk Systems.** With respect to other relatively mature systems of similar (~1 Gyr) age, η Corvi is approximately 1000 times too bright at 24 um. Why? Our recent spectral modeling studies of other bright, dust-rich circumstellar systems can help guide our reasoning and narrow the possibilities.

Observations of the HD113766 high IR excess system (Lisse *et al.* 2008) found a huge amount of dust, at least a Mars' mass worth, closely resembling the inferred makeup of the common S-type asteroids, and composed of a "normal" mix of olivine and pyroxene and metal sulfides. The dust appears to be created by impact and collisional processes at low velocities in a dense primordial asteroid belt most likely undergoing planetary accretion – there is < 1% abundance of silica in the dust. While the compositional mix of η Corvi and HD113766 do have some general similarities in the silicate components (Fig. 2), the warm η Corvi circumstellar dust has much, much more amorphous carbon and strong features in the 6-8 μm range indicative of C-rich species, silica, water ice, and possibly even ferrous sulphate to be derived from S-type asteroidal material. A similar contrast is found between the η Corvi and circumstellar material and the hyper-dusty debris disk found in the solar aged system BD+20-307 (Weinberger *et al.* 2011).

Study of the HD172555 circumstellar material (Lisse *et al.* 2009) finds a high preponderance of fine amorphous silica species (> 45% by surface area, 67% by Si atom count), SiO gas, and large pieces of blackbody dust (impact rubble), at least a Pluto's worth of dust mass, and a very non-solar composition, indicative of a hypervelocity impact between rocky bodies. 4 more young "silica systems" (HD 23514, HD 145263, HD 15407A, and HD 131488; Rhee *et al.* 2008, Lisse *et al.* 2010, Melis *et al.* 2010) have been identified by our group via spectral modeling as containing appreciable silica, confirming their importance. While η Corvi 's dust does have amorphous silica in its makeup, it is not the single most dominant species, nor is there any evidence for SiO gas due to volatilized rock. There are clear silicate emission features at 20 - 35 μm, so that large blackbody grains are not very abundant. We do have to explain the presence of abundant, but not dominant, amorphous silica in the compositional mix. An impact at moderate velocity, at 5 - 10 km/sec, of a primitive Kuiper Belt object, containing a dispersed mixture of silicate dust in an icy matrix, onto a rocky body would provide enough specific energy per kg near the point of impact to melt and vaporize silicates, but would not efficiently convert all silicates in either body; higher relative impact speeds of >10 km/sec are required to do this (Lisse *et al.* 2009 and references therein). A moderate 5 – 10 km/sec range of relative impact velocities would rule out direct transmission of a body on a highly eccentric, non-aligned orbit from the η Corvi Kuiper Belt to ~3 AU in the THZ ($v_{impact}$ ~ 50 km/sec; Wyatt *et al.* 2010). It also rules out any impacts onto a body larger than the Earth ($v_{escape}$ = 11 km/sec). We note, however, that the upper velocity bound for this interaction might increase for very high relative abundances of water and water ice in the Kuiper Belt object, as they can act as thermal buffers absorbing the impact energy (J. Melosh, private comm., 2010).

Examination of the 12 pc distant, 2 - 5 Gyr old HD69830 system shows a high preponderance of olivine rich dust, and no evidence for volatile material of any type, a moderate amount of total mass, and a decidedly non-solar elemental abundance, implying a rocky, processed asteroidal source for the material (Lisse *et al.* 2007b, Beichman *et al.* 2005, 2011). η Corvi 's circumstellar matter has about equal proportions olivine, pyroxene, and silica, abundant carbon-rich material, at least a 200 km radius asteroid's mass, and roughly solar composition refractories, implying a very different, much larger parent body than for the HD69830 system. Similar to the measurements of the HD69830 system, observations of the distant, ~100 Myr old ID8 and P1121 systems (Gorlova *et al.* 2011) find a large amount of dust, about the same as detected around η Corvi, but composed of a mix of olivine and pyroxene, amorphous carbon, phyllosilicates and metal oxide species, again as expected for debris created by disrupting a C-type asteroid (30-40% of all asteroids in the solar system main belt) or shock heating carbonaceous chondrite meteoritic material (Tomioka *et al.* 2007). C-type asteroid material, while relatively primitive vs. S-type asteroid matter, has still undergone volatile depletion and extensive aqueous alteration compared to cometary or primordial nebula material. The ID8/P1121 material is silica-poor, < 1% abundance, indicating dust release by impact and collisional processes at low velocities. η Corvi does not have the amorphous pyroxene, phyllosilicates, or iron oxides of these debris disk systems, while evincing much more amorphous carbon, water ice, and silica, and strong features in the 6-8 μm range indicative of carbon-rich species.

As mentioned above, the flux of the η Corvi excess most closely matches the spectra of primitive water and carbon-rich material found in solar system KBOs and the ~10 Myr old HD100546 circumstellar disk material (Fig. 2b). The η Corvi warm dust also shows evidence for much more silica, another tracer of impact processing, with ~50% of all Si atoms in silicas. While somewhat similar in spectral signature to the primitive material seen released from comets by ISO and *Spitzer* (Lisse *et al.* 2006, 2007b, Reach *et al.* 2010, Sitko *et al.* 2011), including features in the 6 to 8 μm range due to water, PAHs, and organics, these features are much stronger in η Corvi and HD100546 than in comets; comets also do not show evidence for silica, down to the < 5% abundance level (Lisse *et al.* 2007b, Sargent *et al.* 2009). Solar system KBO material is carbon-rich (Barucci *et al.* 2005, Dalla Ore *et al.* 2009) and impact processed (Farinella & Davis 1996, Stern 1996, Brown *et al.* 2010, Brown 2010). On the other hand, η Corvi's flux is somewhat weaker in the PAH lines than HD100546's (although this is most likely an excitation effect; see Lisse *et al.* 2007a) and the η Corvi dust shows evidence for less amorphous silicate and phyllosilicate material, suggesting a drier, and potentially collisionally dominated, processing history for the KBOs forming at ~150 AU from the F2V primary than the HD100546 circumstellar dust at ~13 AU from its A0V primary.

**5.3.1 Parent Body Candidates.** The parent body of the dust detected around η Corvi is a bit of a puzzle, but we have some clues. Without any other evidence, the age of the system alone would lend us to infer from analogous events in our own solar system that the production mechanism for the new dust is either a stochastic random collision between asteroidal bodies, a massive comet outburst, or impacts during an LHB event. From the presence of strong 5 to 8 μm emission features, the near-solar elemental abundance, and the approximately 1:1 ratio of olivine to pyroxene dust, ***we can quickly rule out an asteroidal parent body*** (Fig. 2, 7); the observed dust is too primitive and carbon- and ice-rich to have been differentiated and processed much, as the material in most asteroids has been.

We can also rule out the case of small parent cometary bodies. Ruling out a cometary parent body is somewhat harder to do than for asteroids; while the spectral match is relatively poor (Fig. 2), the η Corvi dust is similarly primitive (Fig. 5) in the majority silicates, which tend to dominate the mid-IR emission spectra. ***But in fact it is too primitive for cometary dust***, in that it is super organics and carbonaceous species rich, having retained many species that small cometary bodies cannot gravitationally bind [e.g., the deviation of HCNO species from solar abundance in comets (Lisse *et al.* 2007a) and CI chondrites like Orgueil (Lodders 2003).] The mass of warm dust observed, at least 9 x $10^{18}$ kg, is 3 to 4 orders of magnitude more than found in even the largest solar system comet presently known, C/1995 O1 Hale-Bopp.

***The most likely storage mechanisms we know for large amounts of such primitive material in the solar system would be inside a large (≥ 100 km radius) icy body kept at low temperature throughout its lifetime in the outer solar system – i.e., a KBO.*** The similarity between the HD100546 and the KBO-derived η Corvi spectra (Fig. 2) are then no accident – both contain large amounts of primitive material found in primordial stellar disks near the end of the disk lifetimes at 5 - 10 Myr age. The mass of warm dust observed, at least 9 x $10^{18}$ kg, is very comfortably about the same amount of mass as found in the larger Centaur bodies, smaller Kuiper Belt objects, or the smaller sized moons of the outer giant planets.

More clues are available from the cold dust disk present in the η Corvi system and current models of Kuiper Belt evolution. Assuming a collisional cascade in this system starting with planetesimals a few km in size, Wyatt *et al.* 2005 infer from the system age of ~1 Gyr a total mass of $20M_{Earth}$ in planetesimals in the belt at 150 AU. This is similar to the mass inferred for the collisional cascade of the disk around the young Herbig Ae star HD100546 of age ~10 Myr (van Den Acker *et al.* 1997, Weinberger *et al.* 2003), and for Fomalhaut at age ~200 Myr (Wyatt & Dent 2002; Table 3). If we compare to our own solar system, this is also similar to the estimated primordial mass of the Kuiper Belt, but on the order of 2 magnitudes larger than inferred for its present day mass. The presence of vast amounts of cold Kuiper Belt dust in η Corvi at ~1 Gyr, corresponding to ~0.04 $M_{Earth}$, argues for recent collisional formation of the dust at ~150 AU. Such continuous dust production will be more efficient in a dynamically excited and perturbed Kuiper Belt, where collisions will be both more frequent and more violent (i.e., dust producing). In such a perturbed belt, the scattering of objects into the inner regions of the system would also be frequent. To produce the observed warm dust in the η Corvi system, all that is required is one object on the order of $10^{-6}$ $M_{Earth}$ be scattered into the inner few AU of the system ~$10^3$ yrs ago where it was disrupted by a collision with a planetary sized object. The same collision, if it occurred at moderate velocities, would transform a partial fraction of silicates into silicas (Fig 4), create a non-solar elemental abundance pattern (Fig. 5), and an olivine:pyroxene ratio off the normal trend line found for dusty disk systems (Fig. 7). (We should point out, however, that while we favor 1 large impact event as the simplest possible explanation and have written this paper as such, the impact of multiple smaller KBO bodies, of total equivalent mass greater than or equal to a 170 km radius, 0.4 g $cm^{-3}$ density body, could also have caused what we observe, as long as the bodies are large enough (≥ 10 km) to produce long-lived km-sized ice-rich fragments and abundant μm- to mm-sized silica dust.)

**5.3.2 Ureilites – Local Evidence for KBO Impacts.** ***The closest spectral match we have found for solar system material to the Spitzer η Corvi dust excess spectrum is the rare Ureilite meteorite type*** – and the match is quite good (Fig. 4f, where we compare the silica and water ice removed *Spitzer* spectrum to the average transmission spectrum measured for the Ureilite meteorite fall of 2008 at Almahata Sitta in the Sudan (Sandford *et al.* 2010; Fig. 4). Ureilites contain the Mg-rich olivine, Ca-poor pyroxene, metal sulfides, carbon-rich matrix, nano-diamonds (in fact, we were led to look at the Ureilites as a match because of their high nano-diamond content), and a little of the silica we find in the η Corvi dust. These meteorites are thought to be derived from an ultra-primitive, carbon rich precursor that has undergone a complex history of shock, disruption, collisional grinding, and re-accretion: ***"The textures, mineralogy, and depleted trace element composition of Ureilites suggest an origin in a partially melted (basalt-depleted) asteroidal mantle of a carbon-rich protoplanet, the size of which is debated (Wilson et al. 2008). At the end of the igneous period, when some melt was still present, the UPB [Ureilite Parent Body] experienced a giant collision that shattered the mantle into 10–100 m-sized pieces (Herrin et al. 2010) and extracted the rest of the melt rapidly…Most workers agree that after the giant collision, the fragments of the UPB reassembled into a jumbled state, possibly around the remnant of the original body. That body was subsequently hit to produce a family of daughter asteroids, a process that was repeated more than once over the history of the solar system, given the collisional evolution needed to get from approximately 10 m-sized UPB fragments to some of the finer grained clasts found…Based on the presence of nonureilitic material in Almahata Sitta, material originating from different parent bodies must have become mixed in."*** (After Jenniskens *et al.* 2010; Fig. 8.) Studies of Ureilite meteorites consistently show evidence for both a primitive silicate mineral phase rich in carbonaceous materials mixed with differentiated and dry igneous phases (Goodrich 1992 and references therein) as well as formation under significant $CO/CO_2$ partial pressure, exactly what we would expect for a moderate to high (5-10 km/sec) impact between an icy, carbon-rich outer solar system primitive body and a rocky, dry inner planetesimal/planet.

Thus there is good evidence in our solar system for material like we observe orbiting around η Corvi, likely produced by the collision between two bodies, a KBO and an inner system rocky body. As such, it should be sourced by inputs from each of these bodies. The ~60% of dust in the silicate and organics-rich portion is derived from original fragments and post-impact re-accretion of relatively unaltered material from each of these bodies. The similarity between the η Corvi and the ~10 Myr old HD100546 excess emissivity spectra argues that the bulk of this unaltered portion of the η Corvi dust (> 75%) is derived from the incoming KBO. Whether this material was shed prior, during, or as a result of the collision is not clear, although its localization near 3 AU implies that it could not have been shed much outside of this distance. Ureilite studies argue, though, that some (the minority, we estimate ≤ 20%; see Fig. 4f) of the meteoritic silicates & carbonaceous material, and thus by inference, ≤ 20% of the η Corvi silicates & carbonaceous dust, comes from the impactee.

By contrast, the ~30% (by surface area; 50% by molar weight of Si) of the η Corvi dust which is in the form of silica material must be formed in an impact at high temperatures and velocities. Silica is not found in abundance in comets, primitive HD100546 disk material (Fig. 4f and Lisse et al. 2007b), Ureilite meteorites, or even in other early solar system materials. Since it is not primordial but is instead produced in high velocity impacts, we believe it is scarce because silica is not easily incorporated into any re-accreted post-impact bodies; formed at high velocity it escapes re-accretion into the local gravity wells and is instead blown out of the system by radiation pressure or absorbed into the star as fine dust after orbital P-R drag decay (see §4.5). Does it come from the impactor or impactee? Preliminary modeling of the many possible impact scenarios and outcomes suggests that the large majority of material ejected in a 2-body collision at km/sec interaction velocities derives from the impactor if the masses of the two bodies are in the ratio $M_{impactee}/M_{impactor} > 100$ or more. Thus the silicates ***and*** silica should be impactor-derived in the most likely cases, i.e., for an impact between a medium sized KBO on the order of 200 km in radius (or many, many smaller KBOs of equivalent total mass) and a planetesimal or planet of radius 1000 km or greater (O'Keefe & Ahrens 1982, 1985; Melosh & Tonks 1993; Benz *et al.* 2007; Marinova *et al.* 2011, given the > 9 x $10^{18}$ kg of dust mass present found in this work, and assuming ≤ 20% efficiency of ejecting material from the impacts between 2 large bodies (O'Keefe & Ahrens 1982; Benz *et al.* 2007; Marinova *et al.* 2011).

In the more improbable other limiting case of an ~200 km radius KBO impacting a roughly equivalent ~200 km radius asteroid orbiting η Corvi, it is possible that more, up to 50%, of the warm, ~3 AU circumstellar material was sourced by the impactee. This fraction should be considered a ***very*** conservative upper limit as it is based on models of impacts assuming both bodies were of equal size and strength, no impactee atmosphere was involved, and we ignore the role of water ice buffering. [As the impactor becomes weaker, the impactee atmosphere more dense, or the impactor more ice-rich, the fraction of ejecta from the impactor rises.] E.g., a typically water rich KBO impactor (Water Ice/Silicates ~ 1) will lose much of its collisional excitation energy via water evaporation, rather than silicate transformation, essentially buffering and protecting its silicates. It will also be more prone to fragmentation, breakup, and emission of silicate and water-ice rich ejecta. The water-poor impactee would still absorb and channel the received impact energy via silica production, so that the observed silica in the equal sized body impact case would derive mainly from the impactee. (N.B., extensive water ice thermal buffering could also mean that the relative velocity of interaction of the two bodies could be higher, up to 30% greater than the 5-10 km/sec quoted for the partial conversion of dry silicates to silicas, while still allowing impactor silicate survival and making the creation of silicas from the impactee even more efficient.)

We can thus argue from simple impact models and first principles that the η Corvi silicate/carbonaceous debris is likely impactor composition dominated in all scenarios, and we recover the close match between the η Corvi and HD100546 silicates; the main source of the silicas, however, could be from either body and is ill determined. However, caution suggests that that detailed thermo-physical and compositional impact modeling, beyond the scope of this paper, is required to properly address these issues and study all possible impact scenarios.

**5.4 Eliminated Model Scenarios.** In our study of the η Corvi system, we have considered a number of non-large impact means of creation and dispersal of the observed material, but we can rule out each in turn on simple physical grounds. Three other main mechanisms come to mind: cometary sublimation, volcanic/geyser emission, and P-R drag of Kuiper Belt dust. Sublimation from a cometary body can produce primitive circumstellar material, but of somewhat dissimilar composition (Figure 2, §2.3) and even the largest observed solar system comet would produce orders of magnitude less dust than the observed η Corvi excess.

Volcanic or geyser emission from a primitive composition icy moon (e.g., Enceladus (Waite *et al.* 2009) or Phoebe (Clark *et al.* 2006)) driven by gravitational tides or an orbital resonance as it orbits around a gas giant parent could possibly produce the right mix of material, as suggested by recent *in situ* Cassini measurements of water and silica rich dust emanating from the interior of Saturn's moon Enceladus (Postberg *et al.* 2009). However, there are no known volcanoes on any of the solar system planets or moons, even Io, that could erupt material into interplanetary space. The same strong gravitational tidal heating that drives these sources also binds the material released from the small source moons into giant planet system orbits. Further complicating this scenario is that such a moon would be the equivalent of an active Io or Enceladus-like body around a warm Jupiter at 1.5 AU in the solar system, with LTE ~ 250K, and would most likely consist of highly processed and differentiated material.

Another possible creation mechanism for the warm dust seen in the η Corvi systems is the forced in-spiraling, via P-R drag, of dust produced in the system's Kuiper Belt at ~150 AU, in a continuous disk of material extending down to the inner reaches of the system. In order to be consistent with our finding of the hottest, smallest dust grains at 350 K, this in-spiraling must stop at ~3 AU from the primary (§4.3). Secession of the in-spiraling could be caused by sublimative losses of the icy component of Kuiper Belt dust, leading to a "sublimation barrier" (Kobayashi *et al.* 2008, 2009) and eventual repulsion and out-sweeping of the dust by radiation pressure. [In our own solar system, with an ice line at ~2.5 AU, the first bodies with notably stable surface ice reside at 5.2 AU and beyond, and HD100546 has an extensive disk of cold dust massing > $10^{22}$ kg reaching from its Kuiper Belt to its ice line at ~13 AU (Grady *et al.* 2005, Lisse *et al.* 2007a).] The expected minimum distance for this effect to operate, though, is significantly outside the system's ice-line, which for η Corvi at $L_\star \sim 4.9\ L_\odot$ is at ~5.5 AU.

Another potential method of terminating the dust in-spiraling at ~3 AU would be the presence of a large Jupiter-sized body at this distance, much as models of the solar system's Kuiper Belt material (Liou and Zook 1999, Stark and Kuchner 2009) show a huge drop in the inflowing material at the orbit of Saturn. As in the collisionally produced case, this argues for a planet-sized object at 3 AU and a large production rate of Kuiper Belt dust, perhaps produced by a giant planet at 94 to 115 AU (i.e. in a 2:1 or 3:2 resonance with 150 AU KBOs). Arguing strongly against the in-spiraling scenario, however, is the new work of Smith *et al.* (2009) who used VLT interferometry to localize the η Corvi warm dust to locations within 0.16 – 2.98 AU, and that of Bryden *et al.* (2009), who used *Spitzer* MIPS 70 μm photometry, CSO/SHARCII 350 μm photometric imagery, and 8-13 μm Keck Interferometer Nuller visibility measurements to demonstrate that there is no significant amount of dust in the regions between the cold outer submm disk and the inner warm dust reservoir, and that the two dust populations are distinctly separated in space. Further, no large Jupiter-mass planet at ~3 AU has been reported for the η Corvi system (Lagrange *et al.* 2009). We thus conclude that P-R drag is not the dominant mechanism at work in this system's outer disk; collisions and blowout will dominate if the dust density is high.

**5.5 Collisional Implications, Planet Detections, & LHB Delivery.** Combined with the independent sub-mm observation of a large amount of cold icy KBO material located at ~150 AU in η Corvi (Wyatt *et al.* 2005, Matthews *et al.* 2010), indicating a massive, excited Kuiper Belt producing ~0.04 $M_{Earth}$ of dust by collisions (the LTE ~ 35K dust is much too cold to be a sublimation product), our spectral and mineralogical analysis argues strongly for warm dust at ~3 AU that is not only derived from a KBO or KBOs, but from bodies scattered into the inner regions of the η Corvi system within the last few thousand years by the same mechanism that caused the large sub-mm excess at ~150 AU (Wyatt *et al.* 2005, 2009; Moor *et al.* 2010; Fig. 8). (Note that the observed collisional debris was created via gradual in-spiraling of a scattered KBO due to multiple scattering and collisional grinding events (Bonsor & Wyatt 2012), and not direct scattering into a collision (Fig. 8) - with perihelion velocities with respect to the η Corvi primary star on the order of 50 km/sec, any body scattered onto an orbit with aphelion at 150 AU & perihelion at 3 AU would have so much specific kinetic energy relative to the ~20 km/sec of a planetesimal in an approximately circular orbit at 3 AU to melt, vaporize, & transform all the usual olivine & pyroxene constituents into silica hypervelocity impact products.) A planetary migration scenario (Gomes *et al.* 2005, Wyatt et al. 2005) would provide the Kuiper Belt excitation and scattering required to produce all the observed phenomena in the η Corvi system – as Neptune migrated outward in our solar system, it swept mean motion resonances across the inner KB objects, pumping up their eccentricities scattering objects throughout the solar system, forcing eventual collisions. Raymond *et al.* (2011) argue instead that the simple presence of a large outer planet, coupled with a massive Kuiper Belt, can lead to LHB-instabilities and scattering on Gyr timescales. In either case, the estimated "transfer time" for such a gradual in-scattering in our solar system is on the order of $10^3$ to $10^5$ yrs (Morbidelli 2011, priv. commun.), of the same order as the estimated age of the η Corvi dust belts. It is exciting to consider that the same series of circumstances could have led to the formation of the polyglot, compound, primitive and carbon-rich parent body for the Ureilite meteorites (Jenniskens *et al.* 2010 and references therein) from a KBO.

Once we accept the KBO collisional hypothesis, another intriguing possibility occurs – that dynamical stirring of the η Corvi Kuiper Belts is ongoing, small icy planetesimals are flying every which way as they are strongly scattered, and that LHB-like collisions are happening - collisions with a small, rocky planet at ~3 AU from the η Corvi F2V primary that is spraying large amounts of free-flying primitive dust and debris into interplanetary space on Kyr to Myr timescales. ***It is important to remember that the detection of collisional products implies not only an impactor, but also an impactee, and are thus an indirect method of planet detection.*** As stated in §4, the mineralogical and size distribution evidence indicates that the impactor and impactee collided more than $10^3$ years ago, at a velocity of interaction between 5 and 10 km/sec. Since the final impact velocity will be the objects' relative velocity, increased by the gravitational acceleration of the impactee (adding roughly an amount $V_{escape}$ to the total), and decreased by any drag forces from the impactee's atmosphere (vanishingly small for an ~100 km radius body, relatively small for a 1 km impactor), the 10 km/s upper limit on the velocity of interaction sets constraints on both the size of the impactee and the orbit of the impactor. Specifically, as relative velocities are likely to be non-negligible, the impactee must have an escape velocity, $V_{escape} < 10$ km/s, and thus a mass significantly below 1 $M_{Earth}$. ***Deep searches for a terrestrial sized planet at ~3 AU, near the middle of the THZ of the η Corvi system, should thus be conducted*** (current searches place relatively weak upper limits of a 2 $M_{Jup}$ sized planet in a 100 day orbit, Lagrange *et al.* 2009). Direct searches will likely be challenging, however, given current capabilities for detection, even for the relatively nearby ***η Corvi*** system at 18 pc distance.

Recent work by Greaves & Wyatt (2010) has suggested that the solar system is a very dust-poor place, when compared to other Sun-like star systems, in the bottom 2-3%, and potentially very planet rich. Both our asteroid and Kuiper belts are thought to be highly depleted, by factors of 100 – 1000, versus their original mass densities (Chambers 2004 and references therein). The natural inference is that the presence of an unusually large number of massive planetary bodies in our solar system has caused the removal of the majority of the potential parent bodies for dust production through planetary migration. The same unusually dust-free solar system we see today underwent an unusual period of small body scattering in the 0.5 – 1.0 Gyr timeframe. η Corvi would seem to be doing the same. (*N.B.* – We estimate that alternative reasons for Kuiper Belt destabilization in η Corvi, like Galactic tides or passing stars, are unlikely to be causing η Corvi's Kuiper Belt excitation. η Corvi is currently at Galactic coordinates l = 296.1792, b = +46.4217 (J2000), farther off the galactic plane than the Solar System by ~√2*18.2 pc, in a relatively low stellar density region, and its Kuiper Belt is unlikely to be affected by galactic or stellar tidal interactions any more than our own solar system is currently.)

This line of reasoning has naturally led us to consider the probable nature of the planet-sized perturbing body at ~100 AU in the η Corvi system. (We note here that Bonsor and Wyatt 2012 argue that a minimum of not 1, but 3 large planetary perturbers, with orbital semi-major axes at approximately 100, 58 and 23 AU, are required to transfer a KBO from 150 AU into 3 AU. Unfortunately from the present set of observations, the dynamical model constraints on the perturbing bodies are not very strong without making some assumptions.) **The main constraints on the putative outermost major orbital perturber are the location of the warm dust, the upper limit of 5 - 10 km/sec on $v_{impact}$, and the location of the perturber.** Strictly speaking, any eccentricity from e = 0 up to e = 1 for the perturber is possible, because the latter would give a collision velocity of 8 km/sec for a KBO and impactee on aligned ($\Delta$Inclination = $0^o$) orbits. We thus end up with a constraint on the location of the outer planet that depends on the assumed mass (or rather escape velocity) of the impactee. We *can* constrain the eccentricity further *if* the impacted object at 3 AU is a planet with an escape velocity of ~5 km/sec in a co-planar orbit, since then the relative velocity of the impactor would have to be less than 5 km/sec, requiring the eccentricity of the impactor and the perturber(s) to be e << 1 in order for the total impact velocity to fall in the range 5-10 km/sec. In this latter case the initial perturber is likely to be an ice giant or Jovian planet in 2:1 or 3:2 resonance with the system's Kuiper Belt objects, i.e., at a distance 94 to 114 AU from the Eta Corvi primary, assuming the Kuiper Belt is centered at 150 AU. Future searches for migrating planetary bodies in this system at 100 – 150 AU from the η Corvi primary are thus highly warranted. They may be difficult, however – as any planet-sized body located in this region will be relatively "cold", within 10-20% of LTE, unlike the hot young Jupiters imaged in the HR 8799 (Marois *et al.* 2009) and Beta Pic systems (Lagrage *et al.* 2009).

Finally, it is interesting to speculate on the effect of the LHB on the Earth and other Earth-like bodies. Timing evidence for an LHB at 4.1 – 3.8 Gya has been found in Apollo samples returned from the Moon (Turner *et al.* 1973, Tera *et al.* 1974) and lunar meteorite isotopic studies (*e.g.* Cohen *et al.* 2000), and in iridium abundance measurements of impact craters in 3.8 Gyr old Greenland strata (Jørgensen *et al.* 2009). It has frequently been noted that the end of the LHB appears to roughly coincide with the first records of life on Earth (Westall 2008 and references therein). It is not clear, however, if the causal connection is due to the delivery of important water and organics-rich material to the Earth by LHB impactors, or due to the clearing out of most of the planet crossing small bodies in the solar system, and the cessation of frequent life-destroying major impacts. ***We can say from this work that the contents of our purported medium-sized Kuiper Belt impactor(s) in the η Corvi system appear to be very water and organics rich, even after impact disruption and fragmentation in the inner system, and appear to have been an important source for delivery of astrobiologically important materials to a rocky planetesimal in the system's THZ, at ~3 AU distance from the F2V η Corvi primary.*** The iridium abundance values found by Jørgensen *et al.* (2009) argue for similarly primitive icy impactors as the cause of the terrestrial Greenland impact craters during the LHB ; the recent fall in 2008 of the Almahata Sitta Ureilite meteorite (Jenniskens *et al.* 2010) provided a direct contemporary example of abundant organics delivery (Sandford *et al.* 2010, Zolensky *et al.* 2010) to the Earth.

Literature estimates of the total amount of mass delivered to THZ objects during the solar system's LHB range from $2x10^{17}$ to the Moon and $4 \times 10^{18}$ to the Earth (Nesvorný *et al.* 2010, assuming multiple comets); $2 \times 10^{18}$ kg to the Moon and $4 \times 10^{19}$ kg to the Earth (H. Levison private comm. 2010, for large KBO deliverers); and $2 - 7 \times 10^{20}$ kg to the Earth of asteroidal meteoritic material (Court & Sephton 2009). These amounts are close to the total amount of warm dust mass we find delivered to ~3 AU in the η Corvi system, $\geq 9 \times 10^{18}$ kg. With respect to water delivery, the mass of circumstellar water ice & gas we detect in the η Corvi system, between $5 \times 10^{17}$ kg (for 5% relative mass of $H_2O$ in $9 \times 10^{18}$ kg total of 0.1 – 100 μm particles) and $5 \times 10^{20}$ kg (for 5% relative mass of $H_2O$ in $9 \times 10^{21}$ kg total in 0.1 – 100 m particles), is on the order of 0.03 to 30% of the mass of the Earth's oceans.

## 6. Conclusions

The circumstellar excess spectrum in the ~1 Gyr old η Corvi system most closely resembles the ~10 Myr old dust found around young Herbig stars with late-stage primordial disks like HD100546, material of near-solar composition that is primitive and still in the process of assembling planetary bodies, accreting onto the central star, and dissipating into the ISM, rather than the material found around HD113766 (S-type asteroid debris) or around the older stars HD69830, ID8, and P1121 (C-type asteroid debris). We interpret this as demonstrating that the parent body for the η Corvi warm circumstellar dust was a large object created early in the system's history, that it formed outside the ice line of the η Corvi system, and that it retained much of its icy volatiles and primitive material. The amount of debris is large, much more than the largest comet ever seen in the solar system (C/1995 O1 Hale-Bopp) and much more like the size of a large Centaur or Kuiper Belt object (r > 130 km), massing enough that if collected into one body, it would mass at least 9 x $10^{18}$ kg (and possibly as much as $10^{22}$ kg if 100m debris fragments exist) and could easily retain abundant primitive carbonaceous material by maintaining a relatively large escape velocity. The ~30% abundance of silica by surface area is also not seen in any of the other cometary systems studied to date by our group, nor is the large amount of 6-8 µm emission due to carbonaceous dusty materials. The dust is in steady state equilibrium with the stellar radiation field and collisional grinding. We conclude that the silica was created at moderate relative velocities (5 to 10 km/sec) in an impact event at ~3 AU created after multiple scatterings and perturbations of a KBO's orbit, an event that occurred between a primitive, water & carbon-rich Kuiper Belt like-object and a small (< 1 $M_{Earth}$) massed planet/planetesimal at ~3 AU over $10^3$ yrs ago. The driver for the infalling KBO as likely a perturbing ice giant or Jovian planet at ~100 AU. ***Thus, we have good evidence at 1.1 – 1.7 Gyr in the*** η Corvi ***system for the impact by a wet planetesimal (or planetesimals of equivalent mass) with radius ≥ 200 km onto a larger body at ~3 AU at moderate relative velocities, delivering significant amounts of water & organic materials to the system's THZ. The timing & relative location of the impact are similar to the predicted events occurring at the Earth and Moon in the solar system's THZ, during the era of the LHB at 0.6 – 0.8 Gyr.*** There are many similarities between this picture and the inferred origin of the organics-rich Ureilite meteorite parent body (Herrin *et al.* 2010), and a Ureilite meteorite fell to the Earth in Sudan in 2008, delivering carbon-rich material. There is also a very good spectral match between the η Corvi *Spitzer* spectrum and the 2008 Sudan Ureilite meteorite spectrum.

## 7. Acknowledgements

This paper was based on observations taken with the NASA *Spitzer* Space Telescope, operated by JPL/CalTech, and on observations taken with the SpeX 0.8-5.5 Micron Medium-Resolution Spectrograph and Imager, funded by the National Science Foundation and NASA and operated by the NASA Infrared Telescope Facility. The authors would like to thank A. Bonsor (bonsor@ast.cam.ac.uk), G. Bryden (Geoffrey.Bryden@jpl.nasa.gov), E. Mamajek (emamajek@ pas.rochester.edu), J. Emery (jemery2@utk.edu), G. Flynn (flynngj@plattsburgh.edu), J. Greaves (jsg5@st-andrews.ac.uk), T. Löhne (tloehne@astro.uni-jena.de), T. Mittal (mittal.tushar22@ gmail.com), D. O'Brien (obrien@psi.edu), J. Plescia (Jeffrey.Plescia@jhuapl.edu), J. Rayner (rayner@irtf.ifa.hawaii.edu), A. Ross (aidan.ross@ ucl.ac.uk), S. Sandford (Scott.A.Sandford@ nasa.gov), C. Stark (christopher.c.stark@nasa.gov), G. Sloane (sloan@isc.astro.cornell.edu), R. Stroud (rms@anvil.nrl.navy.mil), and M. Zolensky (michael.e.zolensky@nasa.gov) for many useful discussions contributing to the analysis and discussion presented in this paper. C. Lisse gratefully acknowledges support for performing the modeling described herein from JPL contract 1274485, the APL Janney Fellowship program, and NSF Grant AST-0908815.

# 9. Tables

**Table 1. Properties of Star η Corvi (HR 4775; HD 109085)**

| Name | Spectral Type | $T_\star$ (°K) | $M_\star$ ($M_\odot$) | $R_\star$ ($R_\odot$) | $L_\star$ ($L_\odot$) | d (pc) | Age (Gyr) | $f_{IR}$/fbol | $r_{dust}$ (AU) | $T_{dust}$ (K) | $v_{dust}$ (km $sec^{-1}$) |
|---|---|---|---|---|---|---|---|---|---|---|---|
| HD10958 | F2V | 6830 - 6950 | 1.4 | 1.6 | 4.9 -5.0 | 18.2 | 1.1 to 1.7 | 3 to 4 x $10^{-4}$ | 150<br>3.0 | 35<br>350 | 3<br>20 |

Data from Nordstrom *et al.* 2004, Chen *et al.* 2006, Matthews *et al.* 2010
For dust @ 3.0 AU and $M_\star = 1.4 M_\odot$, $v = \sqrt{(GM_\star/r)} = \sqrt{(G*1.4M_\odot/3.0AU)} = \sqrt{(GM_\odot/2.1AU)}$

**Table 2. Composition of the Best-Fit Model[a] to the *Spitzer* IRS η Corvi Spectrum**

| Species | Weighted[b] Surface Area | Density (g $cm^{-3}$) | M.W. | $N_{moles}$[c] (relative) | Model $T_{max}$[d] (°K) | Model $\chi^2_\nu$ if not included |
|---|---|---|---|---|---|---|
| **Detections** | | | | | | |
| ***Olivines*** | | | | | | |
| Forsterite ($Mg_2SiO_4$) | 0.30 | 3.2 | 140 | 0.69 | 350 | 11.1 |
| ***Pyroxenes*** | | | | | | |
| AmorphSil/Pyroxene Composition ($MgFeSi_2O_6$) | 0.08 | 3.5 | 232 | 0.12 | 350 | 2.52 |
| Ortho-Pyroxene ($Mg_2Si_2O_6$) | 0.07 | 3.2 | 200 | 0.11 | 350 | 1.62 |
| Diopside ($CaMgSi_2O_6$) | 0.05 | 3.3 | 216 | 0.08 | 350 | 1.32 |
| ***Metal Sulfides*** | | | | | | |
| Ningerite (as $Mg_{10}Fe_{90}S$) | 0.16 | 4.5 | 84 | 0.86 | 350 | 3.21 |
| ***Organics*** | | | | | | |
| Amorph Carbon (C) | 0.14 – 0.20 | 2.5 | 12 | 2.9 - 4.5 | 500 | 3.44 |
| ***Water*** | | | | | | |
| Water Ice ($H_2O$) | 0.15 | 1.0 | 18 | 0.83 | 210 | 2.97 |
| ***Silicas*** | | | | | | |
| Tektite | 0.35 | 2.6 | 62 | 1.5 | 350 | 33.2 |
| (Bediasite, 66% $SiO_2$, 14%$Al_2O_3$, 7% MgO, 6% CaO, 4% FeO, 2% $K_2O$, 1.5% $Na_2O$, Koeberl 1988) | | | | | | |
| **Marginal Detections and Upper Limits[e]** | | | | | | |
| ***Silicas & Silicates*** | | | | | | |
| Quartz ($SiO_2$) | 0.06 | 2.6 | 60 | 0.26 | 350 | 1.07 |
| Fayalite ($Fe_2SiO_4$) | 0.02 | 4.3 | 204 | 0.04 | 350 | 1.04 |
| ***Water*** | | | | | | |
| Water Gas ($H_2O$) | 0.04 | 1.0 | 18 | 0.22 | 210 | 1.03 |
| ***PAHs*** | | | | | | |
| PAH ($C_{10}H_{14}$) | 0.02 | 1.0 | <178> | 0.011 | N/A | 1.03 |
| ***Metal Sulfates*** | | | | | | |
| $FeSO_4$ | 0.02 | 2.9 | 152 | 0.038 | 350 | 1.05 |
| ***Metal Oxides*** | | | | | | |
| Magnetite ($Fe_3O_4$) | 0.02 | 5.2 | 232 | 0.022 | 350 | 1.03 |
| ***Silicates*** | | | | | | |
| AmorphSil/Olivine Composition ($MgFeSiO_4$) | ≤ 0.005 | 3.6 | 172 | ≤ 0.01 | 350 | 1.01 |
| FerroSilite ($Fe_2Si_2O_6$) | ≤ 0.005 | 4.0 | 264 | ≤ 0.01 | 350 | 1.01 |
| ***Silicas, SiO*** | | | | | | |
| Obsidian (75% $SiO_2$, 18% MgO, 6% $Fe_3O_4$) | ≤ 0.005 | 2.6 | 66 | ≤ 0.02 | 350 | 1.01 |
| SiO Gas (100% SiO) | ≤ 0.005 | N/A | 44 | ≤ 0.01 | 350 | 1.01 |
| ***Carbonates*** | | | | | | |
| Magnesite ($MgCO_3$) | ≤ 0.005 | 3.1 | 84 | ≤ 0.02 | 350 | 1.01 |
| Siderite ($FeCO_3$) | ≤ 0.005 | 3.9 | 116 | ≤ 0.02 | 350 | 1.01 |

(a) - Best-fit model $\chi^2_\nu$ = 1.01 with power law particle size distribution dn/da ~ $a^{-3.5}$, 6.3 – 34.7 µm range of fit. 95% C.L. has $\chi^2_\nu$ = 1.06
(b) - Weight of the emissivity spectrum of each dust species required to match the η Corvi emissivity spectrum. No BB fit has been added.
(c) - $N_{moles}$(i) ~ Density(i)/Molecular Weight(i) * Normalized Surface Area (i). Errors are ± 10% (1σ).
(d) - All temperatures are ±10K (1σ). Best fit blackbody function to 6.3 – 34.7 µm IRS spectrum hás $T_{bb}$ = 390 K. LTE @ $r_h$ = 3.0 AU from the 4.9 $L_{solar}$ η Corvi primary = 250 K. The best-fit temperature for the largest water ice grains was found to be 170K.
(e) – Marginal Detections : species which appear to improve the fit by eye, but do not improve the $\chi^2_\nu$ value above the 95% C.L. limit. We call these out in a separate section as while they are not detected by the strict $\chi^2_\nu$ criterion they may be worth pursuing with higher sensitivity and resolution measurements in the future, as are fullerenes, nana-diamonds, and HACs in this system.

## Table 3. Derived Total Masses (in beam) for Dusty Disk Objects Observed by ISO/*Spitzer* and Selected Relevant Solar System Objects

| Object | Observer Distance[1] (pc/AU) | Mean Temp[2] (K) | Equivalent Radius[3] (km) | 19 um Flux[4] (Jy) | Approximate Mass[5] (kg) |
|---|---|---|---|---|---|
| Earth | --- | 282 | 6380 | | 6 x $10^{24}$ |
| Mars | 1.5 AU | 228 | 3400 | | 6 x $10^{23}$ |
| HD113766 (F3/F5) | 130.9 pc | 440 | ≥3000 (300) | 1.85 | ≥ 3 x $10^{23}$ (7 x $10^{20}$) |
| ID8 (G8) | 600.8 pc | 900 | ≥3000 (670) | 0.011 | ≥ 3 x $10^{23}$ (3 x $10^{21}$) |
| Moon | 0.0026 AU | 282 | 1740 | | 7 x $10^{22}$ |
| HD172555 (A5) | 29 pc | 335 | ≥830 (440) | 0.90 | $10^{22}$-$10^{23}$ |
| ***η Corvi (F2) Cold Dust*** | ***18 pc*** | ***~35*** | ***≥ 1600*** | | ***≥ 6 x $10^{22}$*** |
| KBO Pluto | 40 AU | 45 | 1180 | | 1 x $10^{22}$ |
| HD100546 (Be9V) | 103.4 pc | 250/135 | ≥ 910 | 203 | ≥ 1 x $10^{22}$ |
| Asteroid Belt | 0.1 - 5.0 AU | Variable | | | 3 x$10^{21}$ |
| KBO Orcus | 30 - 45 AU | 42-52 | 300 | | 6 x $10^{20}$ |
| Earth's Oceans | --- | 282 | 6380 | | 6 x $10^{20}$ |
| Enceladus | 10.5 AU | 75-135K | 252 | | 1x$10^{20}$ |
| Miranda | 19.2 AU | ~65 | 235 | | 7 x $10^{19}$ |
| ***η Corvi (F2)Warm Dust*** | ***18.2 pc*** | ***400*** | ***≥ 140*** | | ***≥ 9 x $10^{21}$ (9 x $10^{18}$)*** |
| KBO 1996 TO66 | 38 – 48 AU | 41 – 46 | ~200 | | ~4 x $10^{19}$ |
| Centaur Chiron | 9 – 19 AU | 65 – 95 | ~120 | | ~1 x $10^{19}$ |
| HD69830 (K0) | 12.6 pc | 340 | ≥ 60 (30) | 0.11 | ≥ 2 x $10^{18}$ (3x $10^{17}$) |
| Zody Cloud | 0.1 - 4.0 AU | 260 | | | 4 x $10^{16}$ |
| Asteroid | 0.1 - 5.0 AU | Variable | 1 - 500 | | $10^{13}$ - $10^{21}$ |
| Comet nucleus | 0.1 - 10 AU | Variable | 0.1-50 | | $10^{12}$ - $10^{15}$ |
| Hale-Bopp coma | 3.0 AU | 200 | | 144 | 2 x $10^{9}$ |
| Tempel 1 ejecta | 1.51 AU | 340 | | 3.8 | 1 x $10^{6}$ |

(1) - Distance from Observer to Object.
(2) - Mean temperature of thermally emitting surface.
(3) - Equivalent radius of solid body of 2.5 g $cm^{-3}$.
(4) - System or disk averaged flux.
(5) - Lower limits are conservative, assuming dust particles of size of 0.1 - 100 µm (as detected by *Spitzer*, in parentheses), or 0.1µm – 100m (extrapolated using the best-fit power law PSD), ignoring optical thickness effects. For HD172555, we have included the mass of SiO gas & large particle rubble in the estimate. For η Corvi, we quote the directly observed (by remote sensing) 0.1 – 100 µm mass in the abstract and use this quantity for determination of the minimum mass delivered.

# 10. Figures

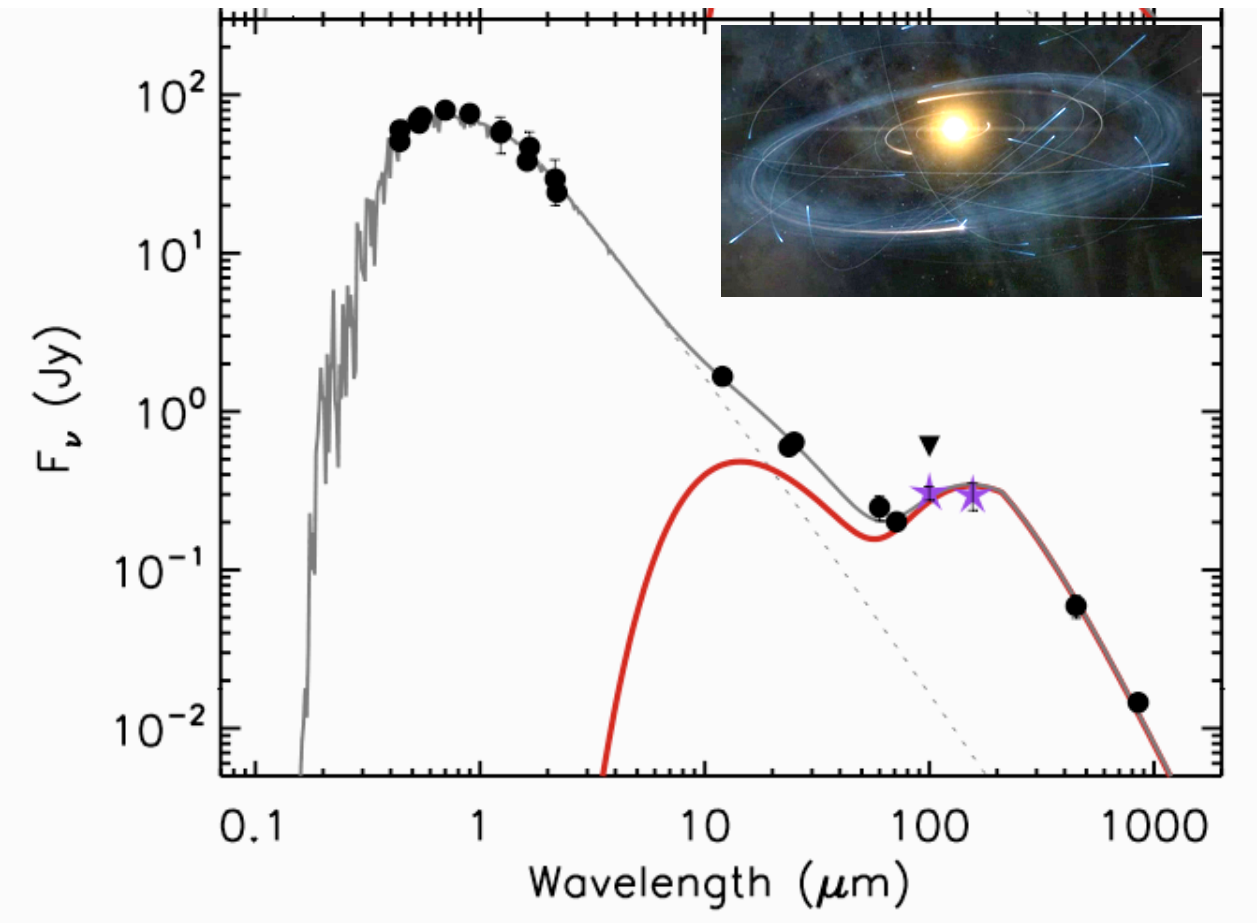

**Figure 1** – ***(a)* SED for η Corvi** showing the 0.4 – 2.2 μm BVR/2MASS system photometry dominated by stellar photospheric emission, the 12 – 100 μm stellar/circumstellar dust MIR flux measurements of IRAS and *Spitzer*, & the 100 – 1000 μm cold FIR excess measured by Herschel and JCMT/SCUBA [2]. Solid grey line: combined fit to the η Corvi SED using a 2-blackbody model (red) with warm (350K) & cold (35K) dust reservoirs + emission from a Kurucz F2V photosphere normalized to the BVR/2MASS photometry (dashed line).

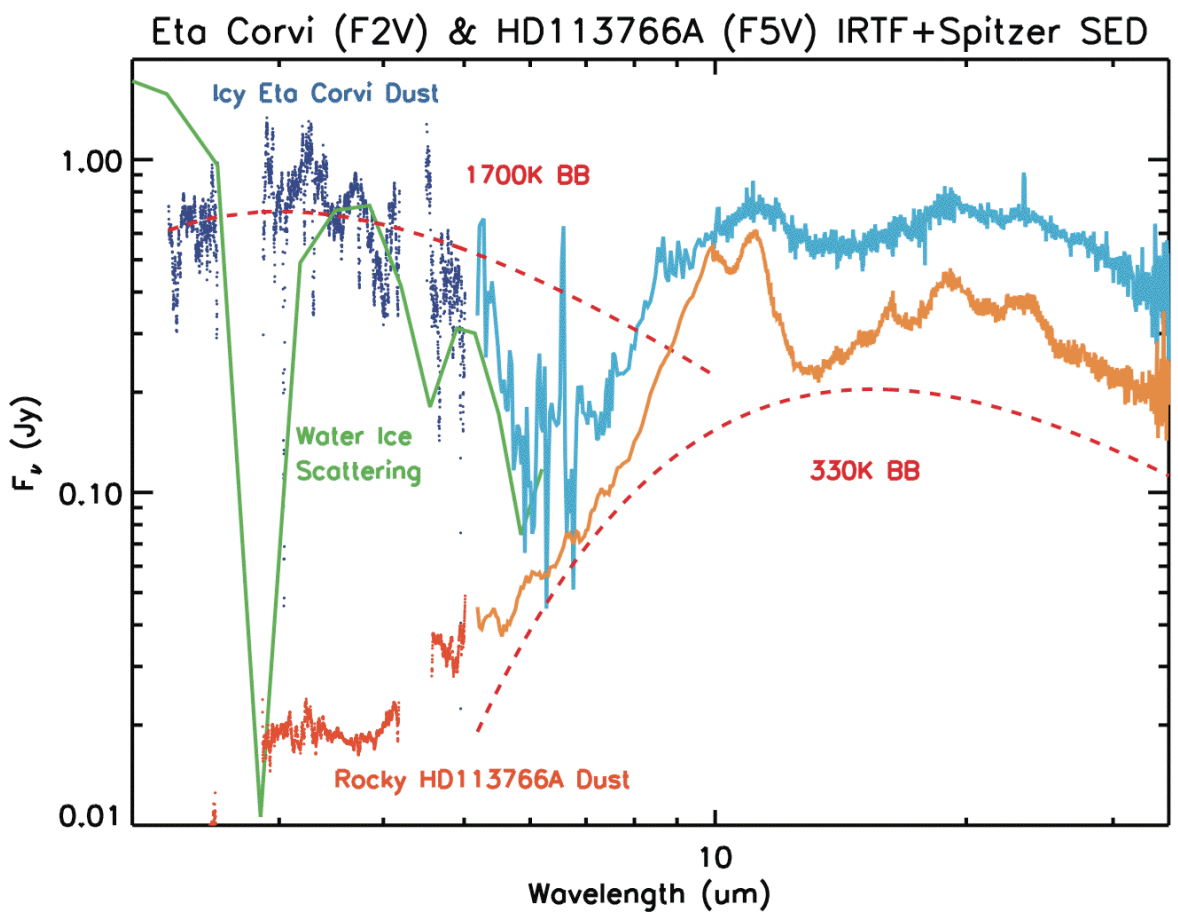

***(b)* Detail of the η Corvi circumstellar excess flux** (blue), as compared to blackbody emission (red) and the rocky circumstellar dust of HD113766A (orange; Lisse *et al.* 2008). The Kurucz model shown in (a) has been subtracted from the *Spitzer* 5-35 μm total spectrum (light blue) and the total SPeX 2-5 μm spectrum (dark blue) in order to determine the excess. The SPeX data corroborate the steep upturn in *Spitzer* flux shortward of 6 μm, and the combined spectrum is consistent with scattering of starlight by high-albedo icy dust (light green). The η Corvi scattered light excess must be from dust inside ~10 AU since the SPeX beam radius is ~6 AU, and since HST did not detect any extension at optical wavelengths beyond ~10 AU (Clampin & Wisniewski 2011, priv. commun.).

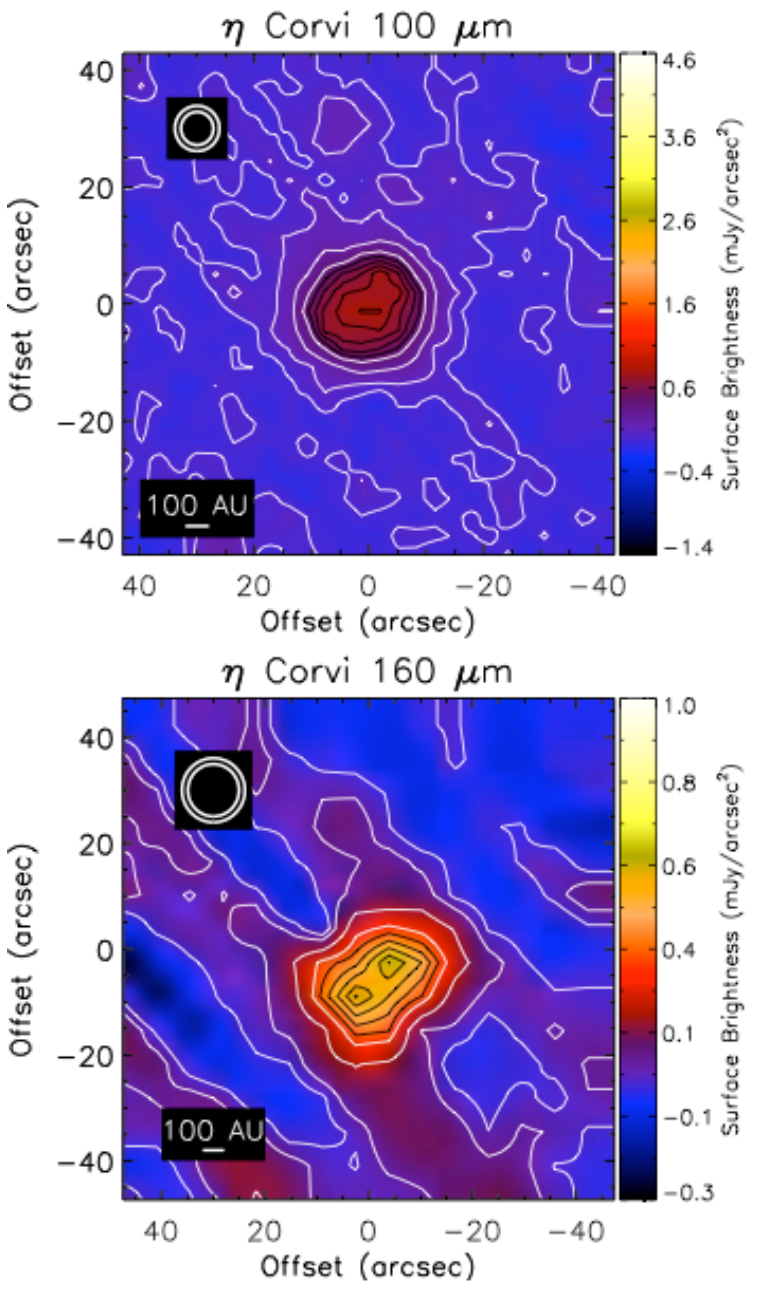

***(c)*** 100 μm (top) and 160 μm (bottom) **Herschel PACS FIR images of η Corvi**, after Matthews *et al.* (2010). Contours are shown at 0, 10, 30, 50, 60, 70, 80, 90 and 99% of the peak in the map. Circles in the upper left corner of each panel mark the nominal beam sizes.

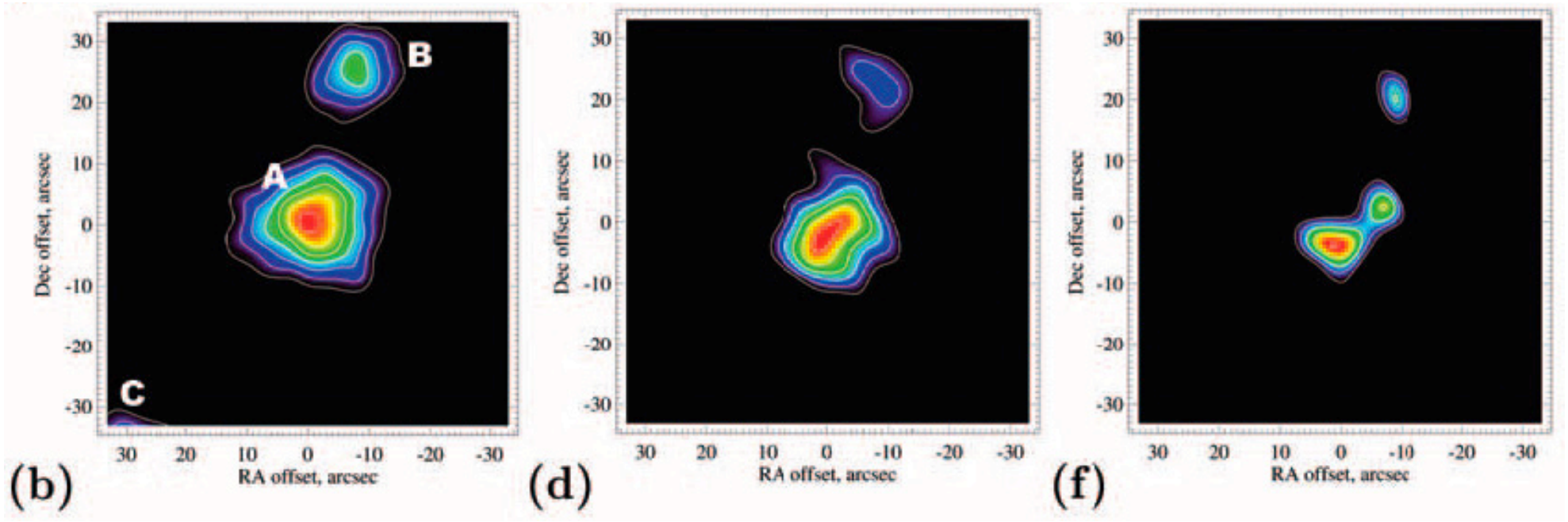

***(d)* JCMT/SCUBA submm images of η Corvi** at 850 μm (15.8” resolution), 450 μm (at 13.7” resolution) and 450 μm with an effective resolution of 9.5” (after Wyatt *et al.* 2005). In all images, the observed sub-mm emission is from cold KB dust at a distance of ~150 AU (LTE ~ 35K) from the primary; the star itself is not visible. Biolabate structure due to a tilted ring system is evident in the 160 μm and high-resolution 450 μm images. The second source toward the top in the sub-mm images is in the background and not relevant.

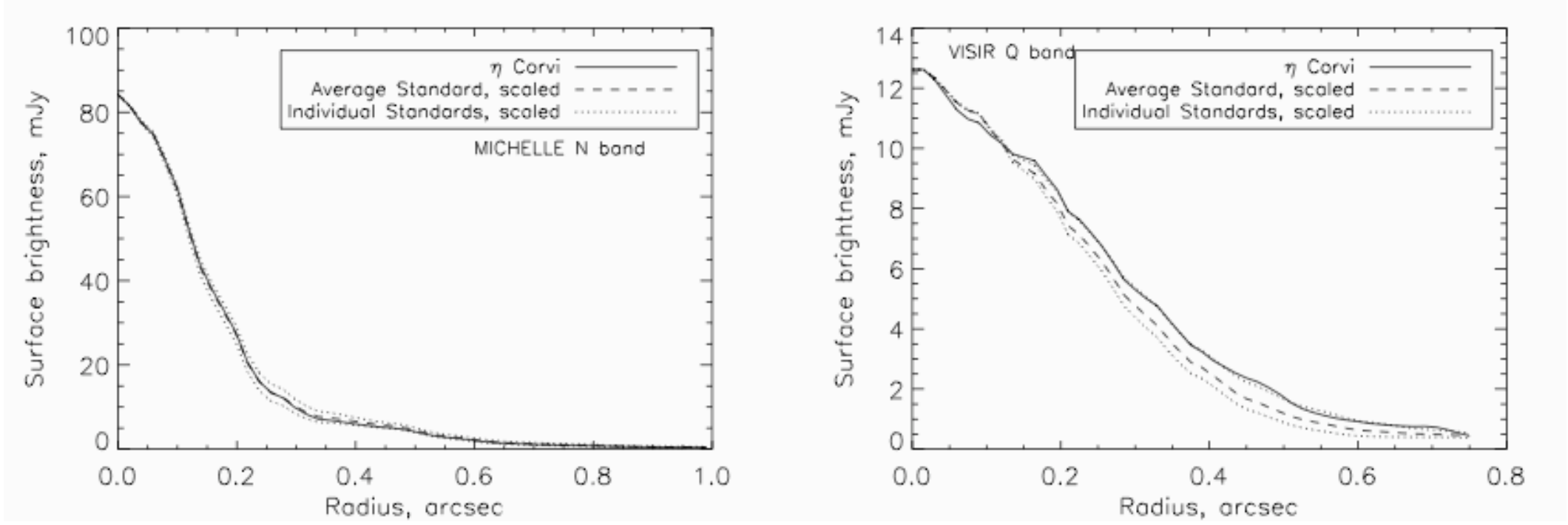

***(e)* GEMINI and VLT profiles of η Corvi** N and Q emission compared to point stellar sources, showing the lack of resolved extension for the warm dust flux outside 3.5 AU. (After Smith *et al.* 2008).

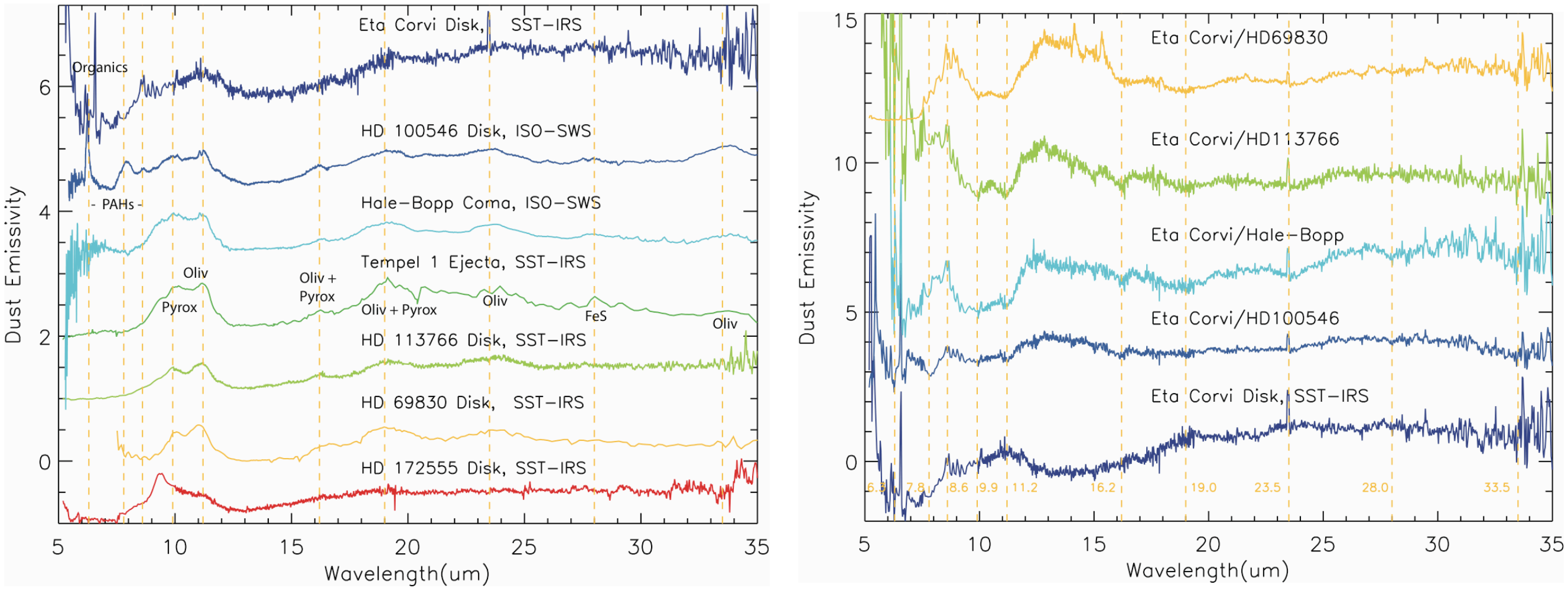

**Figure 2 – (a) Comparison of the mid-IR emissivity spectrum of η Corvi with the spectra of dust** from: a young, organic rich Herbig A0 star building a giant planet (HD100546); two comets (Hale-Bopp and Tempel 1); a young F5 star building a terrestrial planet (HD113766); a mature main sequence star with a dense zodiacal cloud (HD69830); and the silica-rich debris created by a hypervelocity impact in the HD172555 system. Spectra are ordered, starting from the top, by increasing rocky content and processing of the dust. The similarity between the η Corvi and HD100546 spectra is readily apparent. **(b)** Simple ratio comparison of the emissivity spectra of η Corvi, HD100456 (a young Herbig with comets building giant planets), Comet Hale-Bopp, HD 113766 (a young F star building a terrestrial planet from S-type asteroidal dust) and HD69830 (a mature MS K-star with a dense zody from the recent breakup of a C-type asteroid). All 3 of the comparison systems contain young dust, formed in less than 15 Myr. The most primitive dust, found in the disk of HD100546, produces the best match to the η Corvi dust, as can be seen from the relatively small excursions (< ±25%) from the norm in the ratio, predominantly at λ < 16 um. Most of the differences between the η Corvi and HD100546 emissivity can be attributed to differences in the relative abundances of water ice and carbon-rich components (see Fig. 4).

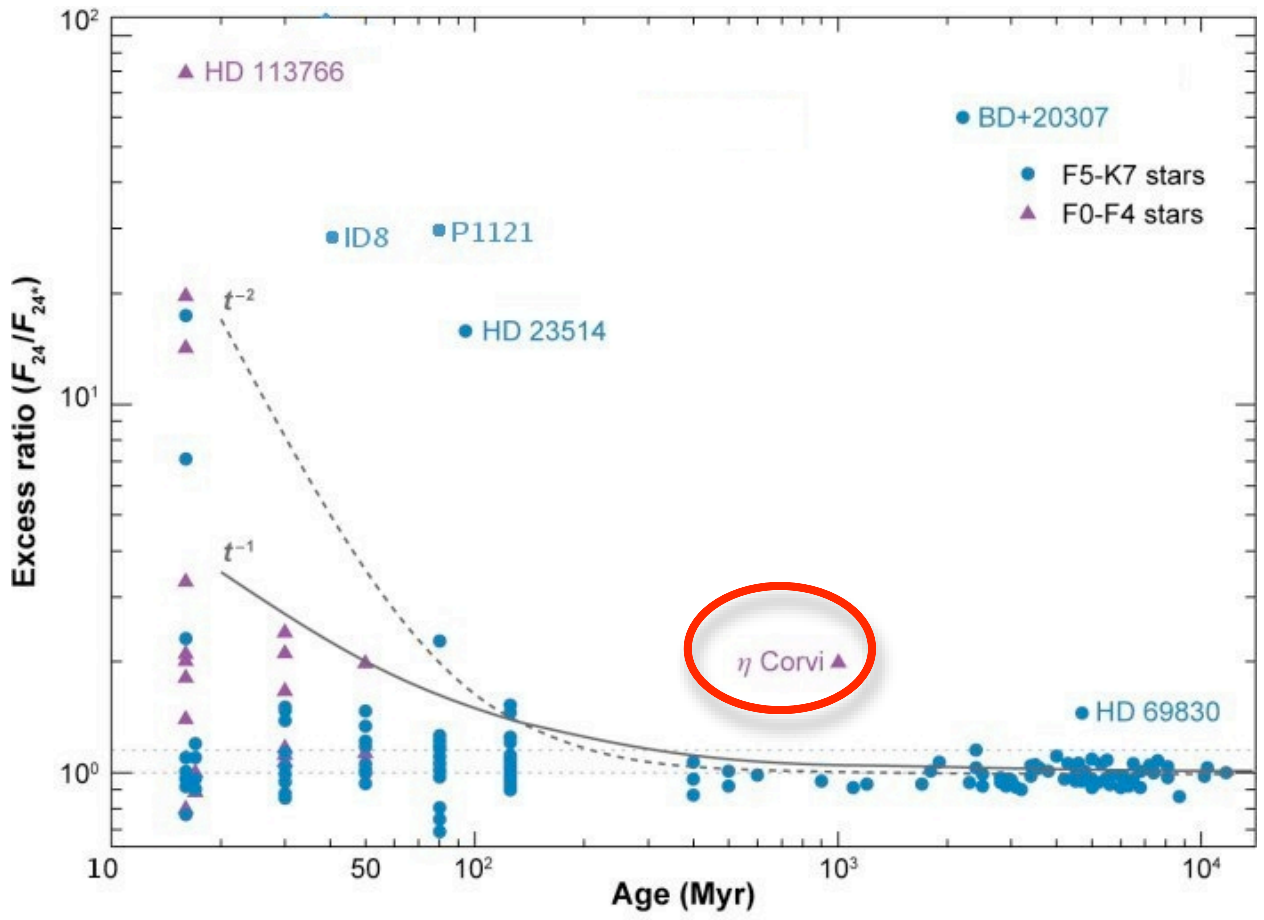

**Figure 3** – **Dusty disk IR excess flux vs. system age.** η Corvi is the 3rd brightest of Chen *et al.* 2006's 59 IRAS-excess systems, and the only one which is a "mature" MS system of ~1.4 Gyr age, or about 1/3 of its total MS lifetime. The 1/t and $1/t^2$ trend lines fit most of the sources in the current sample except outliers like η Corvi, which clearly has a high $L_{IR}/L_* = 3 \times 10^{-4}$ for its age, suggesting something unusual has occurred in this system.

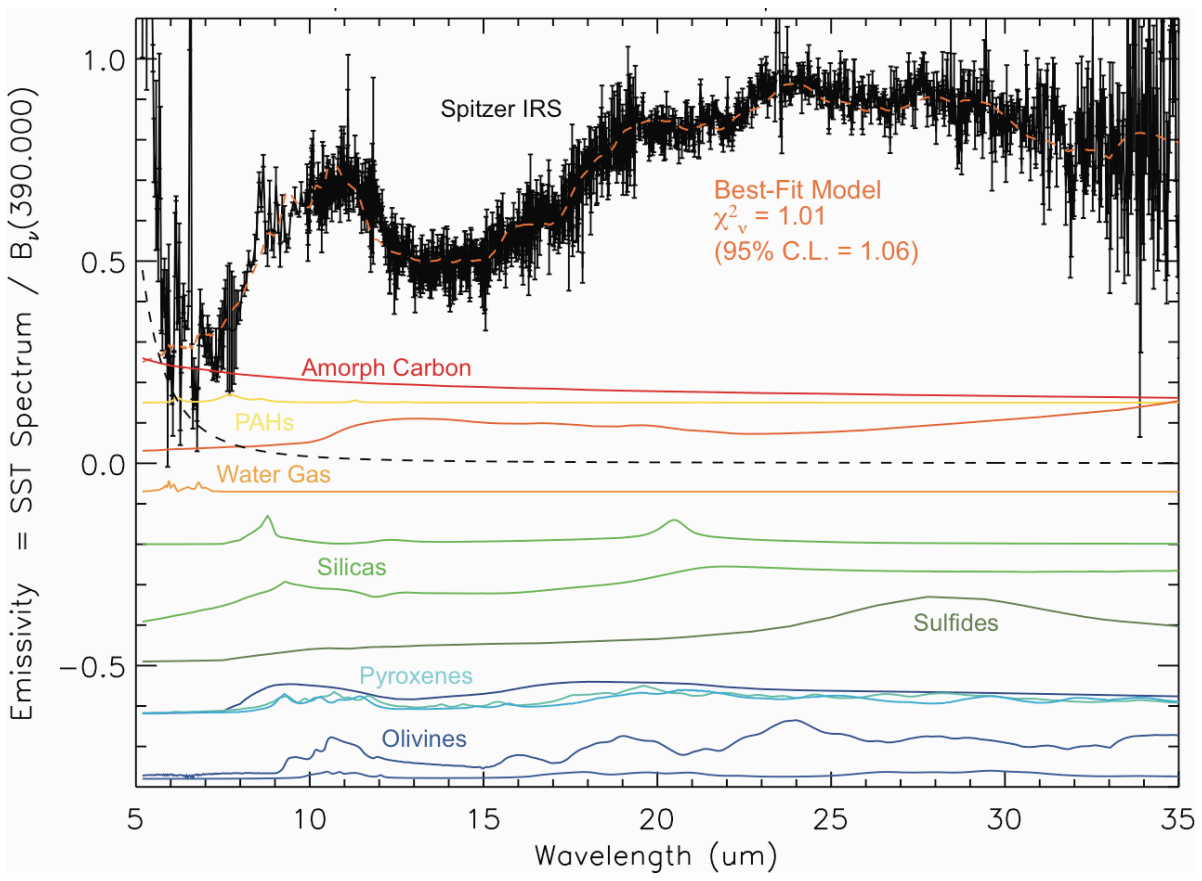

**Figure 4 - *Spitzer* IRS emissivity spectrum of the η Corvi** circumstellar excess (black), and best-fit compositional model (orange dashed line). ***(a)*** Best fit model components using small, solid, optically thin dust grains with ferromagnesian silicates (olivines in dark blue, pyroxenes in purple and light blue) and silica (bright green), amorphous carbon (red), metal sulfide (olive green), water ice (dark orange) & gas (light orange), PAHs (yellow), and amorphous carbon (red). The relative contribution of each species to the total observed flux (Table 2) is given by the amplitude of each emissivity spectrum; for presentation purposes, the large fractions of forsterite and Fe-sulfide have been divided by 2 before plotting.

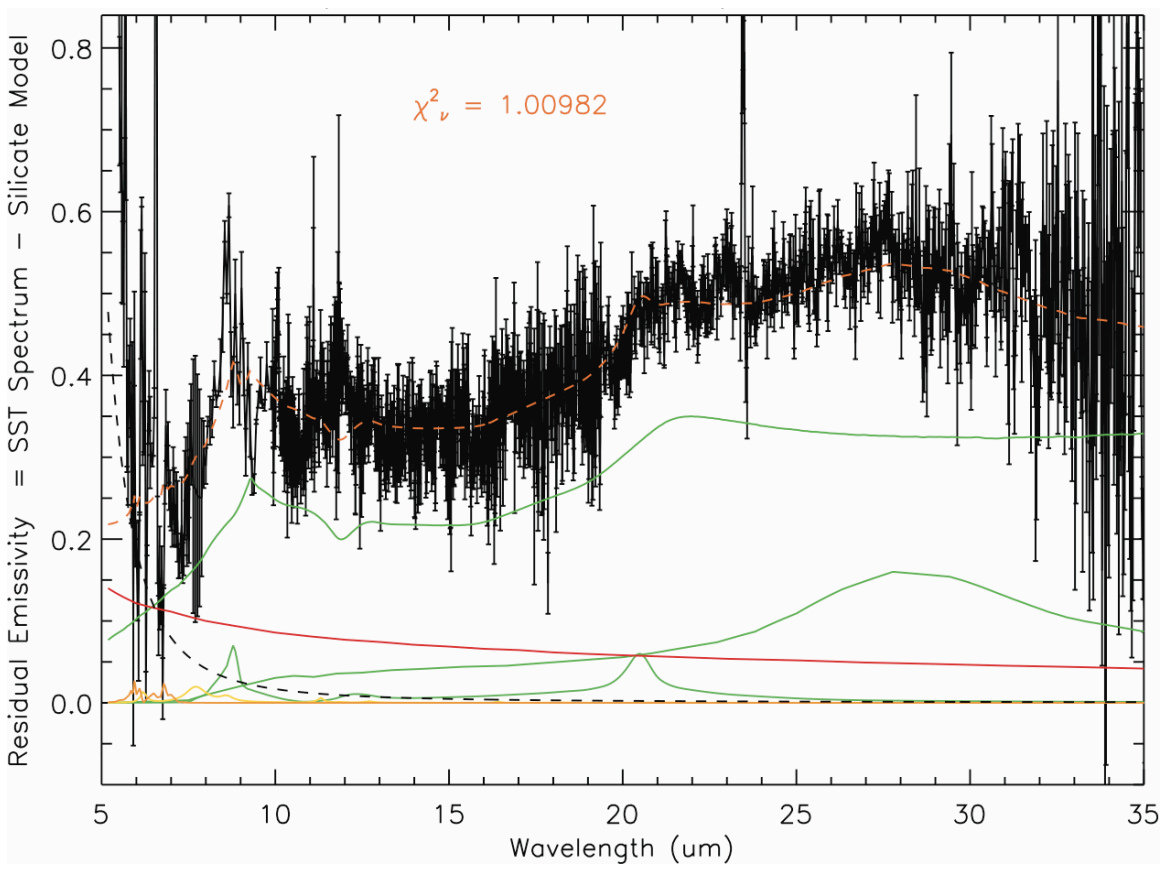

***(b)*** Residuals of the best-fit model to the *Spitzer* IRS emissivity spectrum. All of the usual **silicate** species have been accounted for and subtracted, as well as the water ice contribution to the emissivity. The contributions of the unusual amorphous, glassy silica ('Tektite', broad green and 'Quartz', narrow green spike at 8.5 and 21 um) species can clearly be seen, as can the emissivity contributions of the more typical species amorphous carbon (red), water gas (orange) & metal sulfides (olive green).

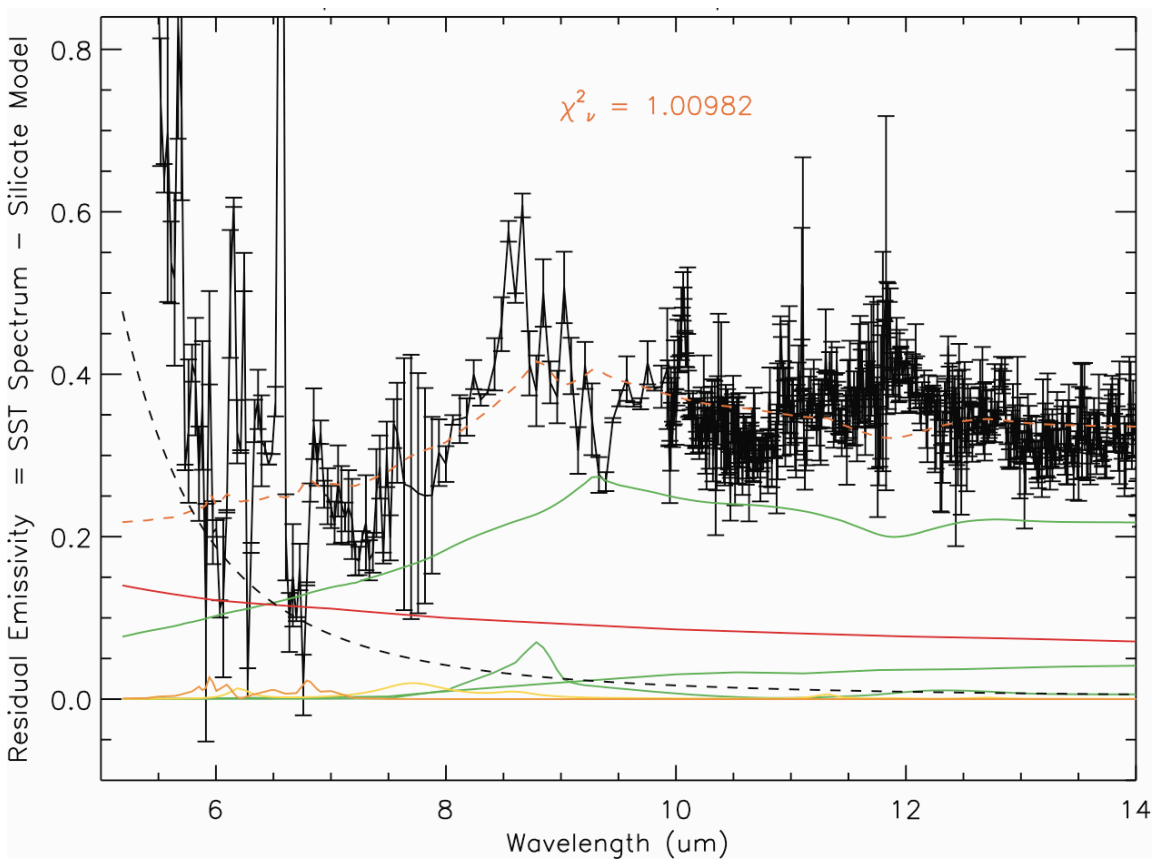

***(c)*** Same residual plot, with the 5-15 μm region emphasized to present details of the poorly fit emission features due to C-rich species at 6-8 μm (likely due to additional nanodiamonds, fullerenes, & volatile Tholins; see Figs. 4d-f).

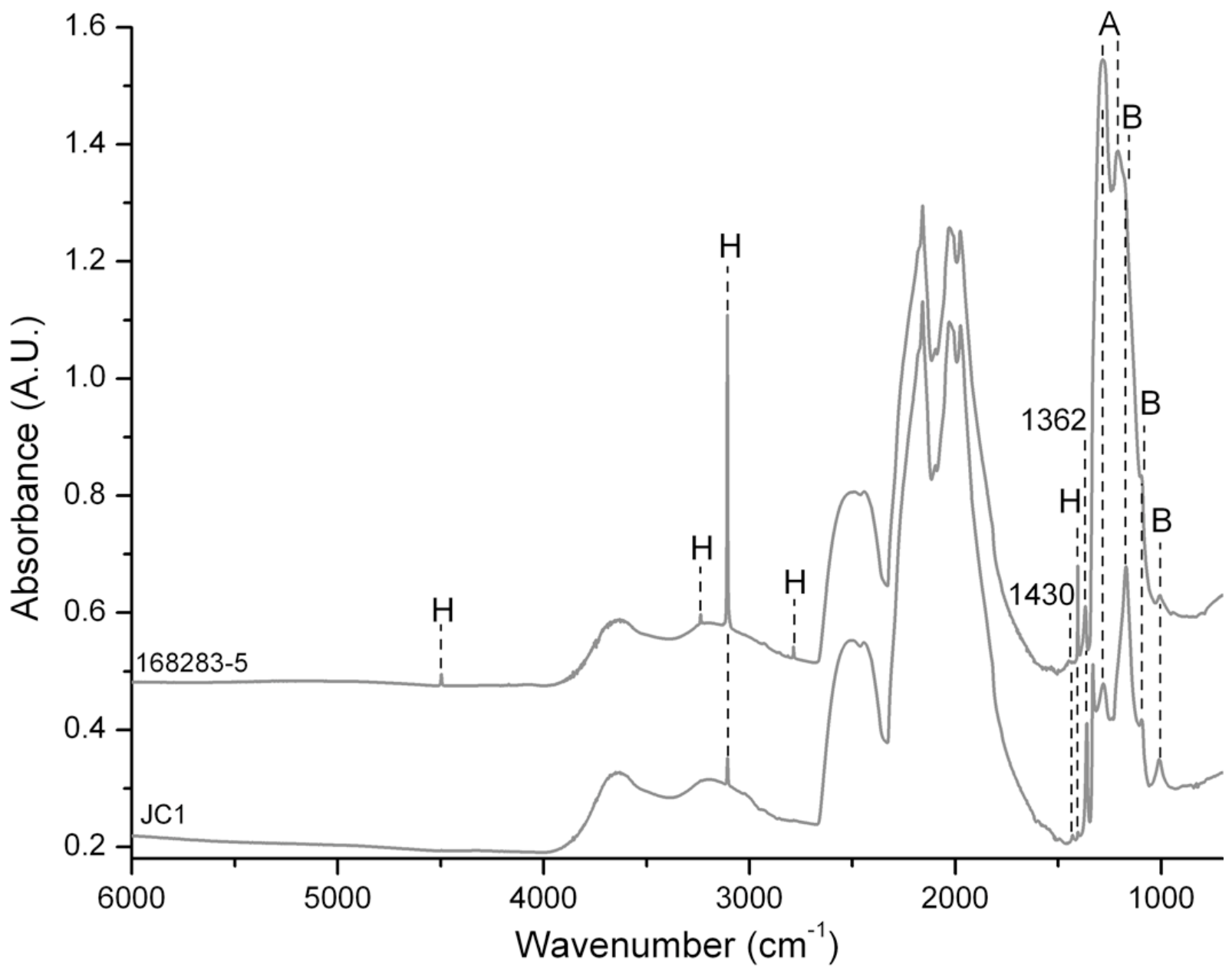

***(d)*** Typical IR absorption spectra of "pink" nanodiamonds from the 2010 study of Gaillou *et al.* The mid-IR behavior is dominated by the main diamond C-C stretch at 1331 $cm^{-1}$, by ppm C-N impurity vibrational features at 1430 $cm^{-1}$, by platelets at 1362 $cm^{-1}$, & by C-H stretches at 1405, 2785, 3107, 3236, and 4496 $cm^{-1}$.

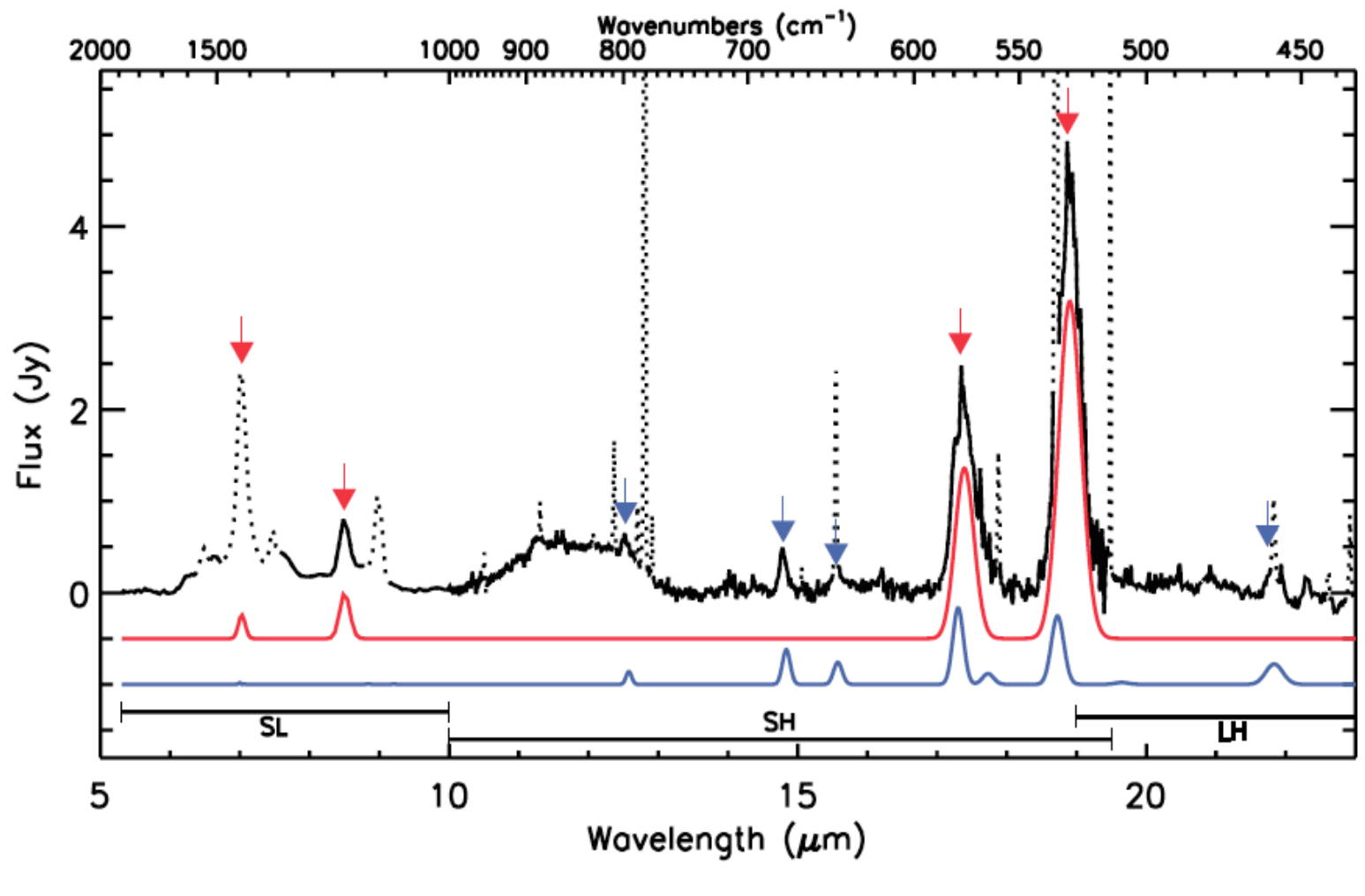

***(e)*** Continuum subtracted and forbidden emission line removed *Spitzer*-IRS spectrum of PN Tc1 from 5 - 23 μm, compared to fullerene emission lines (after Cami *et al.* 2010). Red arrows mark the wavelengths of all IR active modes for $C_{60}$; blue arrows those of the four strongest, isolated $C_{70}$ bands. The red and blue curves below the data are thermal emission models for all infrared active bands of $C_{60}$ and $C_{70}$ at temperatures of 330 K and 180 K, respectively. The broad plateau from 11 - 13 μm is attributed to emission from SiC dust. Apparent weak emission bumps near 14.4, 16.2, 20.5 & 20.9 μm are artifacts. The nature of the weak feature near 22.3 μm is unclear since it appears differently in both nods.

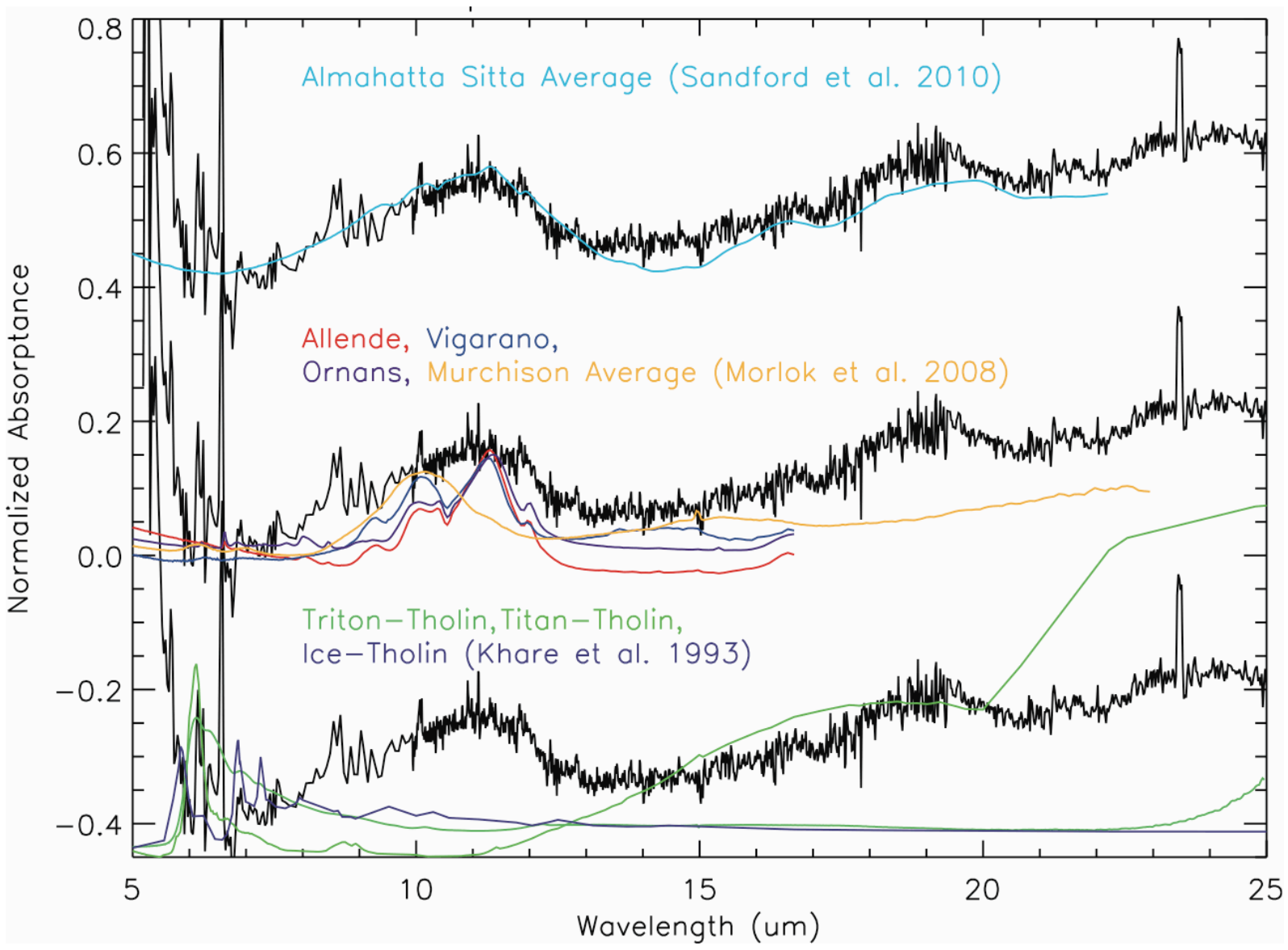

***(f)*** Comparison of the η Corvi carbon/silicate dust emissivity spectrum (i.e., with the altered silica and ephemeral water ice components removed) vs. the transmission spectra of a Ureilite meteorite, four C-rich carbonaceous chondritic meteorites, & three Tholin spectra. ***Above Top***: Versus the Almahata Sitta meteorite fall of Oct 2008. The agreement with the Almahata Sitta spectrum, a Ureilite, is very good, with the exception of small differences in the continuum at 12-15 um (most likely due to uncertainties in our water ice modeling and/or *Spitzer* order matching; a similar problem is seen in the η Corvi/HD100546 spectral ratios, Fig 2), & narrow features at 6-8 μm likely due to additional nanodiamonds, fullerenes, & volatile Tholins. ***Above Middle***: η Corvi dust emissivity spectrum versus transmission spectra from the main primitive, carbon-rich carbonaceous chondrite classes, CI, CM, and CV. The meteorite spectra are poor matches. ***Above Bottom***: Transmission spectra of carbon and ice-rich "Tholins", laboratory materials fabricated from mixes of icy protoplanetary disk materials exposed to UV, x-ray, ionizing radiation, and/or electrical discharges at temperatures and pressures expected for outer solar system small bodies and moons. The absorption features of the Tholins lie in the 6-8 um range, and can account for many of the unexplained features seen in the η Corvi spectrum above the continuum and the Almahata Sitta spectrum; they would also likely be relatively volatile, and thus destroyed easily, before being incorporated into the Almahata-Sitta parent body. ***(g) Below***: Same as above, except comparison of the solar system materials to the ISO HD100546 circumstellar excess spectrum. As for η Corvi, the similarity to Almahata-Sitta is also strong, but mismatches exist, mainly due to the presence of strong PAH features in HD100546 at 6 – 8 μm.

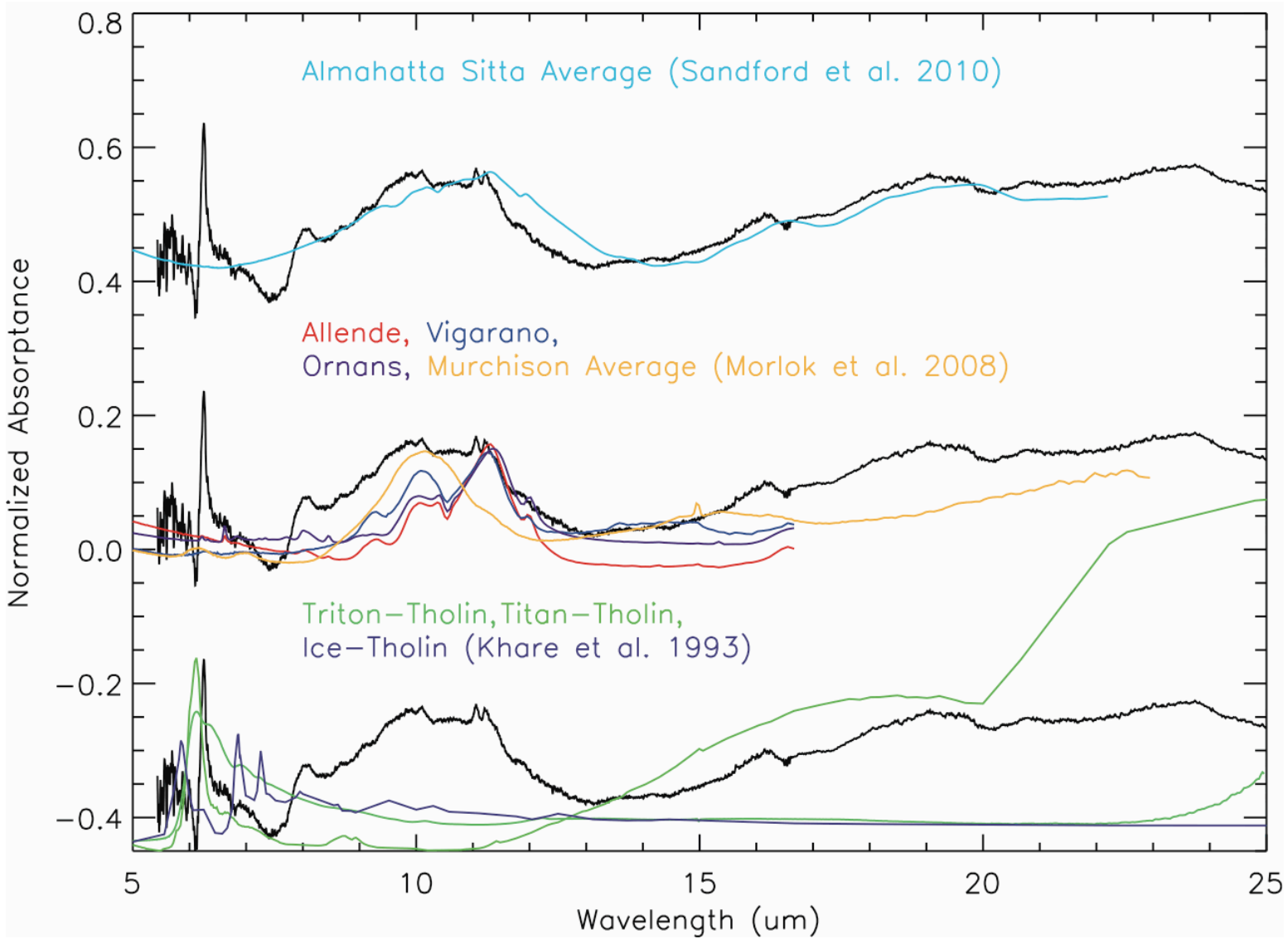

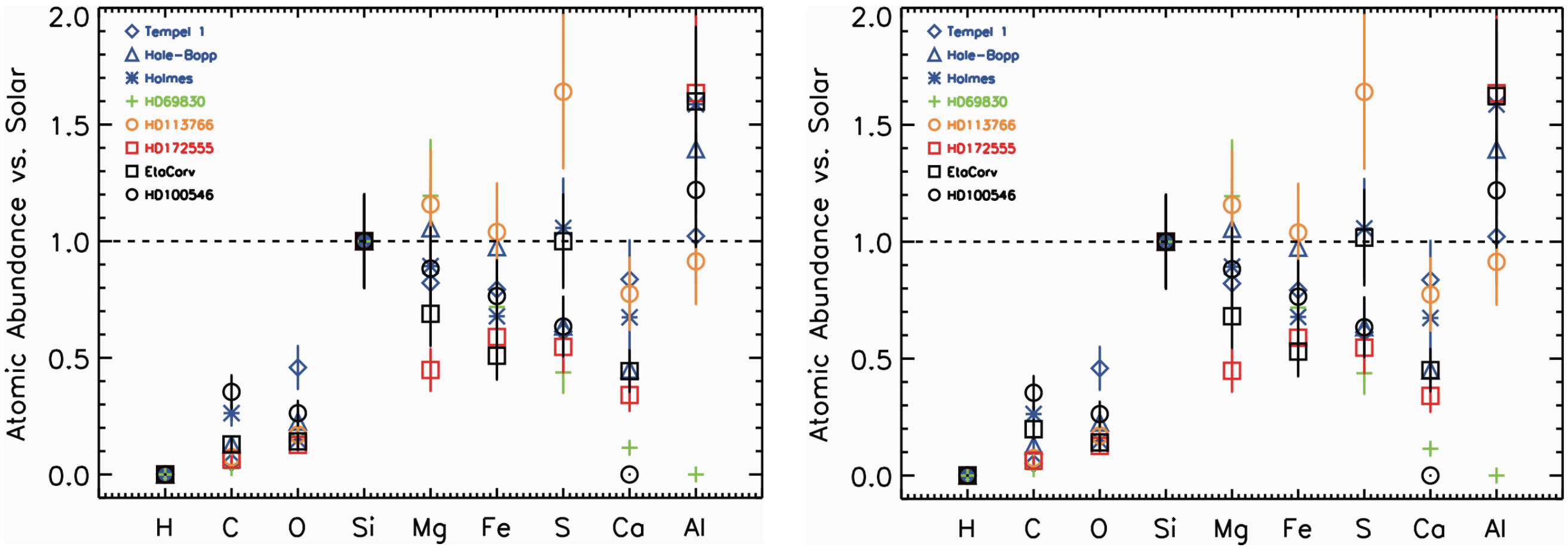

**Figure 5** - Derived elemental abundances for the η Corvi dust excess, relative to solar abundances, and as compared to other *Spitzer* dust spectra (Figure 2). ***(Left)*** High photospheric subtraction, low amorphous carbon model abundances. ***(Right)*** Low photospheric subtraction, high amorphous carbon model abundances. The Si relative abundance has been set = 1.0. The major refractory species, with the exception of S and Al, are all very depleted vs. solar, quite different than what is found cometary dust, which trends near solar. A similar pattern was found for the young debris disk system HD172555, with silica dominated debris formed as a result of a giant hypervelocity impact, and the Herbig disk system, HD100546.

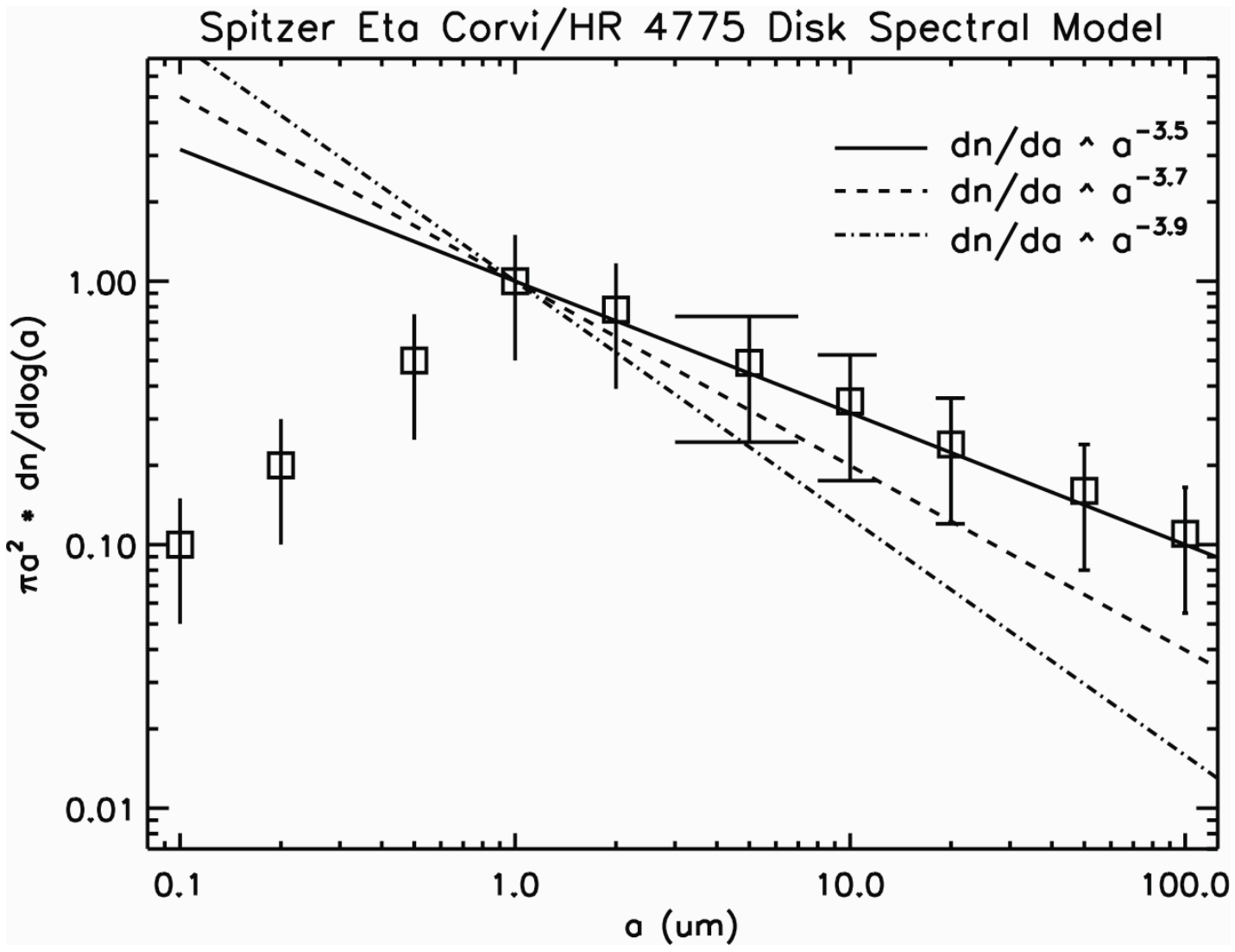

**Figure 6** - Derived particle size distribution for the η Corvi dust excess producing the strong silica and silicate emission features. The derived PSD is close to one in collisional equilibrium with small (< 1 μm) particles removed preferentially by radiation pressure and P-R drag. Error bars are estimated at 50% of the relative abundance at a given size.

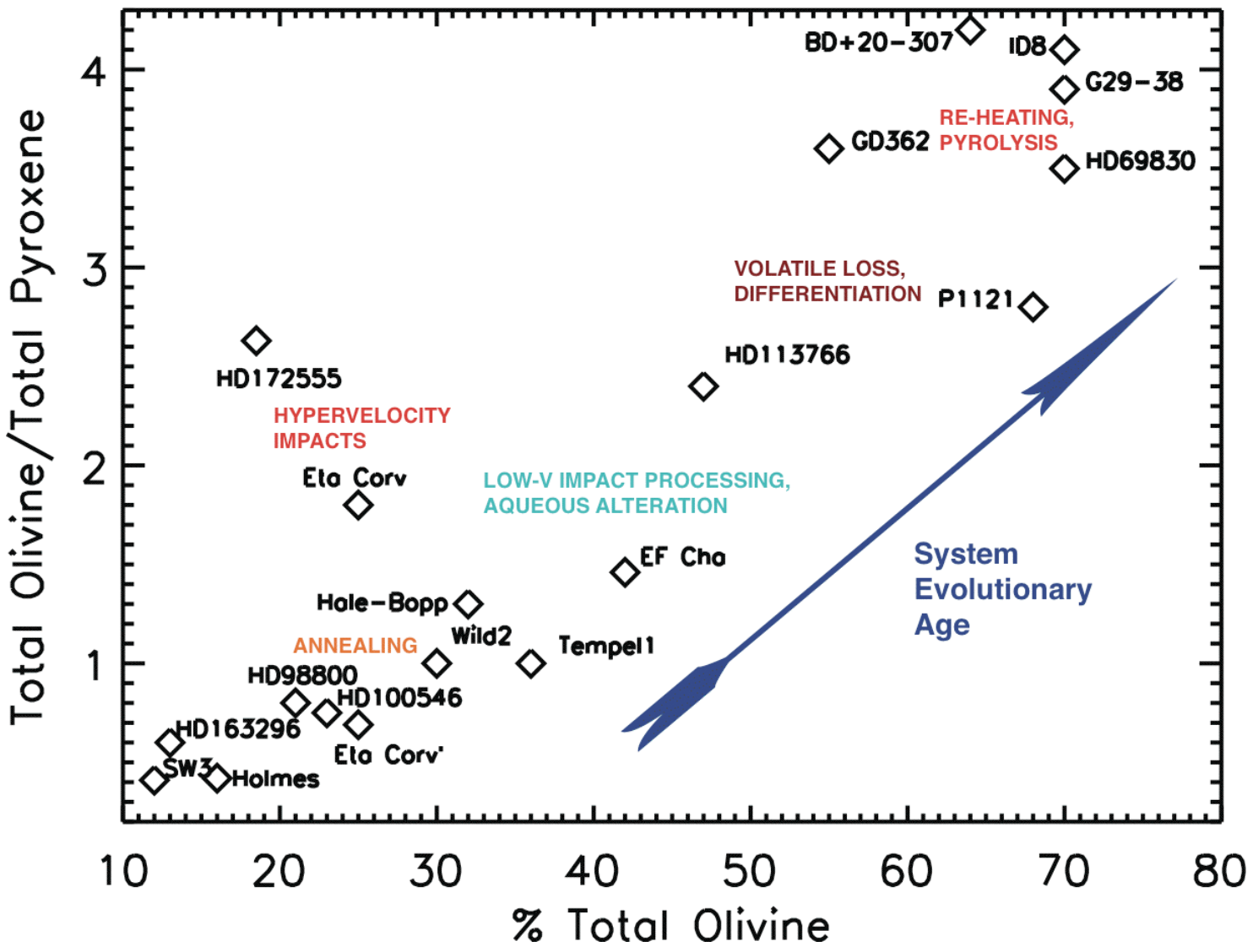

**Figure 7 — Silicate mineralogy for** η Corvi versus that found in 5 comet systems (SW3B/C, Sitko *et al.* 2011; Hale-Bopp, Lisse *et al.* 2007a; Holmes, Reach *et al.* 2010; Tempel 1, Lisse *et al.* 2006; and Wild 2, Zolensky *et al.* 2007, 2008), the primitive YSO disk systems HD100546 (Lisse *et al.* 2007a) and HD163296; the primitive dust found around HD98800B; the crystalline, S-type asteroid material in the ~10 Myr old F5V terrestrial planet forming system HD113766 (Lisse *et al.* 2008); the hypervelocity impact debris created in the 12 Myr HD172555 system; the mature asteroidal debris belt system HD69830 (dominated by P/D outer asteroid dust; Lisse *et al.* 2007b) and the similarly rocky material found in the hyper-dusty, solar aged F-star system BD+20-307 (Weinberger *et al.* 2011); and the ancient debris disk of white dwarfs G29-38 (Reach *et al.* 2008) and GD362 (Jura *et al.* 2007). The general trend observed is that the relative pyroxene content is high for the most primitive material (i.e., YSOs), and low for the most processed (i.e., white dwarfs). Young systems with material altered by high velocity impact processing, like HD172555 and η Corvi, lie above and to the left of the trend line. Note, however, that the η Corvi point is roughly half-way between the aggregate trend line and the location of the strongly altered HD17255 material, consistent with partial transformation of roughly 50% of the η Corvi silicates into silica (see text). Another point (η Corvi') has been included, to demonstrate where the η Corvi material would map to if all the observed silica were derived from the more fragile pyroxene silicate – right next to the location of the best spectral match system HD100546 (Fig. 2), a 10-15 Myr Herbig star with an evaporating and aggregating thick proto-planetary disk of gas and dust. This suggests that the η Corvi parent body must have formed very early, and remained unaltered until a recent high velocity collision released its material.

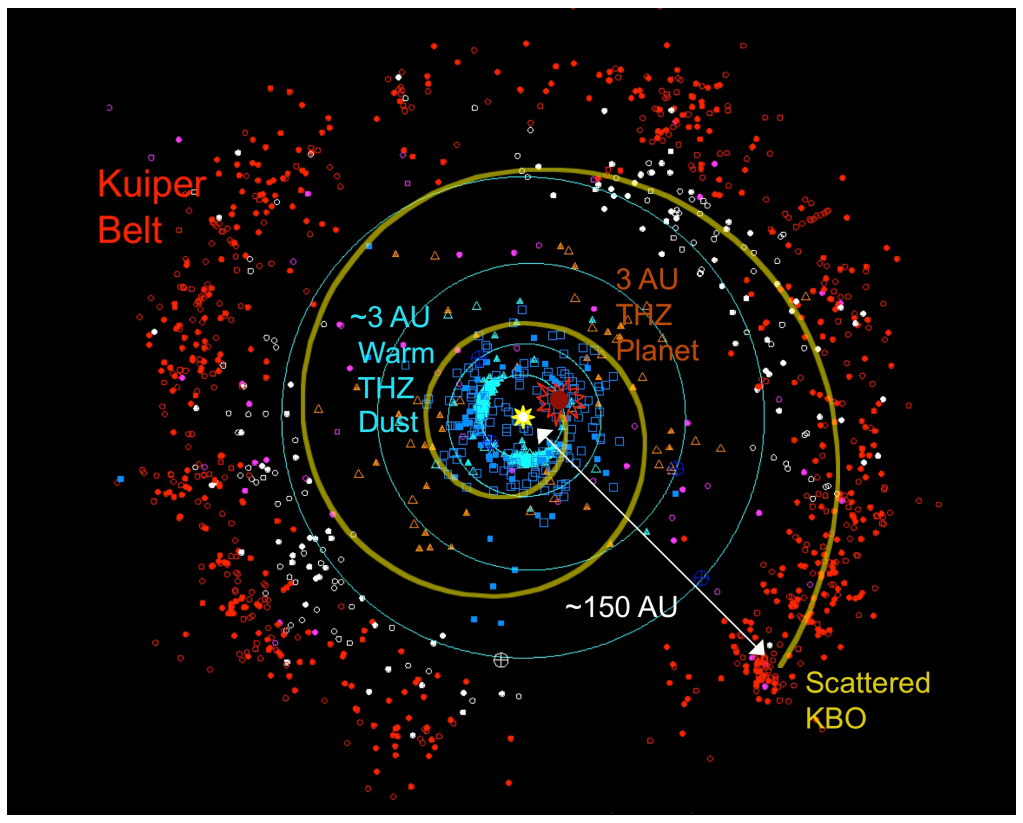

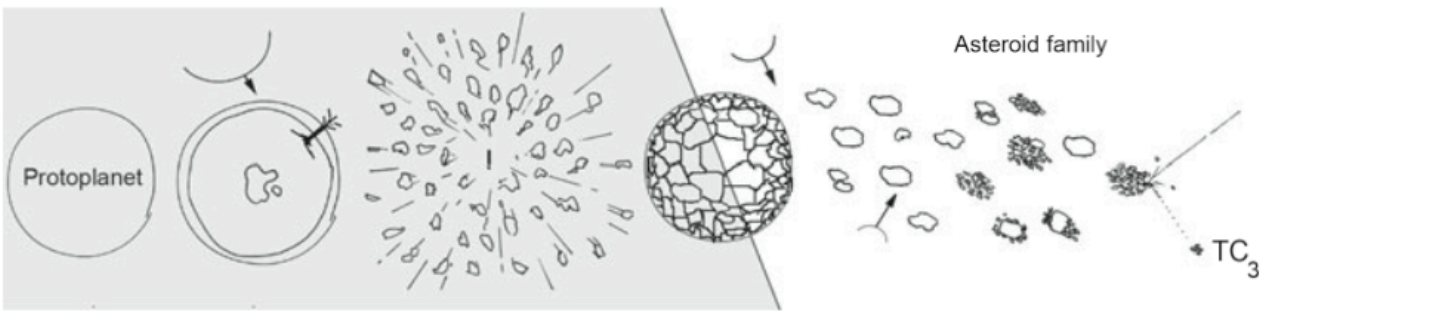

Fig. 1. Cartoon summarizing our understanding of the origin and evolution of the ureilite parent body. From left to right: The ureilite parent body protoplanet underwent partial melting and melt-extraction before a catastrophic disruption occurred that created 10–100 m-sized fragments. The fragments reassembled and migrated to the current position in the asteroid belt; In more recent times, a collision broke the protoplanet and created an asteroid family. The daughter asteroids broke and fragments underwent size reduction to centimeter-size scale in multiple collisions, each time followed by a process of reaccretion into asteroids, during which chondritic material became mixed in with the ureilitic material. More recently, a small asteroid collided with the parent asteroid of 2008 $TC_3$ and released the 3 m object, now exposed to cosmic rays. The small asteroid evolved into a mean-motion resonance and was perturbed into an Earth-crossing orbit, from which it impacted Earth and was recovered.

**Figure 8** – ***Left.*** Schematic representation of the best-fit scenario for producing the observed warm dust in the η Corvi system : Over many Myrs, a medium sized (100 - 200 km radius) KBO was scattered onto an in-spiraling orbit crossing into the inner reaches of the system, where it intersected the orbit of a planetary-sized rocky body located in the THZ, at ~3 AU. Much of the resulting debris was left in near-primary orbit at ~3 AU, where it continues to grind down. The dynamical process causing the KBO scattering continues to stir up collisions in the system's Kuiper Belt, producing cold dust there as well. ***Right.*** Cartoon depicting the estimated history for the Ureilite parent body in the solar system, which is also consistent with the η Corvi KBO – rocky planetesimal impact + subsequent reassembly + fragment collisional grinding scenario. (After Jenniskens *et al.* (2010).)